\documentclass[sn-mathphys-num]{sn-jnl}% Math and Physical Sciences Numbered Reference Style 
\usepackage{graphicx}%
\usepackage{multirow}%
\usepackage{amsmath,amssymb,amsfonts}%
\usepackage{amsthm}%
\usepackage{mathrsfs}%
\usepackage[title]{appendix}%
\usepackage[dvipsnames]{xcolor}%
\usepackage{textcomp}%
\usepackage{manyfoot}%
\usepackage{booktabs}%
\usepackage{algorithm}%
\usepackage{algorithmicx}%
\usepackage{algpseudocode}%
\usepackage{listings}%
\usepackage{hyperref}%

\usepackage{etoolbox}
\usepackage{mdframed}
\usepackage{xspace}
\usepackage{todonotes}
\newcommand{\ang}[1]{\langle {#1} \rangle}
\newcommand{\dang}[1]{\overset{\bullet}{#1}}
\newcommand{\innbr}[1]{\Gamma^-_{#1}}
\newcommand{\outnbr}[1]{\Gamma_{#1}^+}
\newcommand{\mcc}{\text{MCC}}
\newcommand{\deltamcc}{\delta_\text{MCC}}

\newcommand{\fc}{f_C}
\newcommand{\fp}{f_P}
\newcommand{\fb}{f_B}
\newcommand{\fl}{f_L}

\newcommand{\N}{\mathcal{N}}
\newcommand{\M}{\mathcal{M}}
\newcommand{\W}{\mathcal{W}}
\newcommand{\dce}{d_{\text{CE}}}

\newcommand{\A}{\mathcal{A}}
\newcommand{\mcmclong}{\textsc{Maximum Common Network Contraction}\xspace}
\newcommand{\mcmc}{\textsc{MCNC}\xspace}

\newcommand{\setsplit}{\textsc{Set Splitting}\xspace}

\newif\ifcomplete
\completetrue

\newtheorem{thm}{Theorem}
\newtheorem{lemma}[thm]{Lemma}
\newtheorem{prop}[thm]{Proposition}
\newtheorem{corol}[thm]{Corollary}
\newtheorem{remark}{Remark}
\newtheorem{definition}{Definition}

\ifcomplete
\letcs{\dependentprop}{proprep}
\letcs{\enddependentprop}{endproprep}
\else
\letcs{\dependentprop}{prop}
\letcs{\enddependentprop}{endprop}
\fi

\ifcomplete
\letcs{\dependentcorol}{corolrep}
\letcs{\enddependentcorol}{endcorolrep}
\else
\letcs{\dependentcorol}{corol}
\letcs{\enddependentcorol}{endcorol}
\fi

\ifcomplete
\letcs{\dependentthm}{thmrep}
\letcs{\enddependentthm}{endthmrep}
\else
\letcs{\dependentthm}{thm}
\letcs{\enddependentthm}{endthm}
\fi

\begin{document}

\title{Finding Maximum Common Contractions Between Phylogenetic Networks}


%\author{Bertrand Marchand}{Department of Computer Science, University of Sherbrooke, 2500 Boulevard de l¿Université, Sherbrooke, QC J1K 2R1, Canada. }{bertrand.marchand@usherbrooke.ca}{https://orcid.org/0000-0001-8060-6640}{}%TODO mandatory, please use full name; only 1 author per \author macro; first two parameters are mandatory, other parameters can be empty. Please provide at least the name of the affiliation and the country. The full address is optional. Use additional curly braces to indicate the correct name splitting when the last name consists of multiple name parts.
%
%\author{Nadia Tahiri}{Department of Computer Science, University of
%Sherbrooke, 2500 Boulevard de l¿Université, Sherbrooke, QC J1K 2R1, Canada.
%}{Nadia.Tahiri@USherbrooke.ca}{https://orcid.org/0000-0002-1818-208X}{}
%
%\author{Olivier {Tremblay-Savard}}{Department of Computer Science, University of Manitoba, 75 Chancellors Cir, Winnipeg, MB R3T 5V6, Canada.}{olivier.tremblay-savard@umanitoba.ca}{https://orcid.org/0000-0003-2514-0264}{}
%
%\author{Manuel Lafond}{Department of Computer Science, University of
%Sherbrooke, 2500 Boulevard de l¿Université, Sherbrooke, QC J1K 2R1, Canada.
%}{manuel.lafond@usherbrooke.ca}{https://orcid.org/0000-0002-1825-0097}{}

\author[1]{\fnm{Bertrand} \sur{Marchand}}\email{bertrand.marchand@usherbrooke.ca}

\author[1]{\fnm{Nadia} \sur{Tahiri}}\email{nadia.tahiri@usherbrooke.ca}

\author[2]{\fnm{Shohreh} \sur{Golpaigani Fard}}\email{golpaigs@myumanitoba.ca}

\author[2]{\fnm{Olivier} \sur{Tremblay-Savard}}\email{olivier.tremblay-savard@umanitoba.ca}

\author[1]{\fnm{Manuel} \sur{Lafond}}\email{manuel.lafond@usherbrooke.ca}

\affil[1]{\orgdiv{Departement d'Informatique}, \orgname{University of Sherbrooke}, \orgaddress{\street{2500 Boulevard de l'Universite}, \city{Sherbrooke}, \postcode{J1K2R1}, \state{QC}, \country{Canada}}}

\affil[2]{\orgdiv{Department of Computer Science}, \orgname{University of Manitoba}, \orgaddress{\street{75 Chancellors Cir}, \city{Winnipeg}, \postcode{R3T5V6}, \state{MB}, \country{Canada}}}


\abstract{
In this paper, we lay the groundwork on the comparison of phylogenetic networks
based on edge contractions and expansions as edit operations, as originally
proposed by Robinson and Foulds to compare trees.  We prove that these
operations connect the space of all phylogenetic networks on the same set of
leaves, even if we forbid contractions that create cycles.  This allows to
define an operational distance on this space, as the minimum number of
contractions and expansions required to transform one network into another. We
highlight the difference between this distance and the computation of the
\emph{maximum common contraction} between two networks.  Given its ability to
outline a common structure between them, which can provide valuable biological
insights, we study the algorithmic aspects of the latter.  We first prove that
computing a maximum common contraction between two networks is NP-hard, even
when the maximum degree, the size of the common contraction, or the number of
leaves is bounded. We also provide lower bounds to the problem based on the
Exponential-Time Hypothesis.  Nonetheless, we do provide a polynomial-time
algorithm for weakly galled trees, a generalization of galled trees.
}

 %TODO mandatory; please add comma-separated list of keywords



\maketitle

\section{Introduction}

The reconstruction of evolutionary histories, and ultimately of a universal
``Tree of Life'' 
based on biological data is one
of the core tasks of comparative genomics, and bioinformatics as a whole.
However, due to events such as horizontal 
gene transfer~\cite{abby2012lateral,koonin2001horizontal} 
and hybridization~\cite{ellstrand2000hybridization}, 
evolutionary histories may not always be represented as trees.
As a result, the concept of phylogenetic \emph{networks} has emerged
to represent evolution in its full generality,
and has become
a central topic in bioinformatics research~\cite{huson2010phylogenetic}. 
Unlike trees, it allows for the presence of nodes with more than one
parent, usually called \emph{reticulations}.
Reconstructing phylogenetic networks from data is a notoriously difficult
problem, and a wide variety of methods have emerged for tackling
it~\cite{bandelt1999median,boc2012t,camara2016inference,gusfield2004optimal,solis2016inferring}.
However, given the same data-set, different methods may not always yield the same
result, which can be the source of heated debate in the
community (see e.g., the exchanges on the reconstructions of COVID phylogenetic networks~\cite{forster2020phylogenetic, sanchez2020median,forster2020reply,mavian2020sampling}).

This raises the question of the development of \emph{metrics} on
phylogenetic networks, in order to be able to compare different
predictions, evaluate their accuracy against simulated or gold standard data-sets, and identify outliers or similarities among them.
In the case of trees,
one of the most established metrics is the Robinson-Foulds distance~\cite{robinson1981comparison}, due to the simplicity of its definition, its ease of computation,
and the fact that it also yields a maximum \emph{common structure} between the trees as a by-product. 
%Its usual definition is as 
It is usually defined as the size of the symmetric difference of the sets of
\emph{clades} (i.e.\,sets of leaves descending from a single node) of the two
input trees. Note however that its original definition in
\cite{robinson1981comparison} 
%was 
presented it as the minimum number of edge contractions and expansions required to go from one tree to the other.

The situation is not as simple in the case of networks. To start with,
for networks of unbounded degree (i.e.\,allowing for an unbounded number
of ancestors and descendants per node), even deciding whether two
networks are identical is {\sc Graph
Isomorphism}-complete~\cite{cardona2014comparison}. Nonetheless,
several different metrics have been developed for sub-classes of 
networks. Some of them generalize the clade-based definition 
of the Robinson-Foulds metric, such as \emph{hardwired
cluster}~\cite{huson2010phylogenetic,cardona2008metrics},
\emph{softwired cluster}~\cite{huson2010phylogenetic} and 
\emph{tri-partition}~\cite{moret2004phylogenetic,cardona2008metrics} distances.
There exists however even binary networks that are distinct
while exhibiting the same sets of clusters~\cite[Figure
6.28]{huson2010phylogenetic}, and likewise for the 
tripartition distance~\cite{cardona2008tripartitions}. In addition,
in the case of the soft-wired distance, there can be an exponential
number of clusters to compare, making a polynomial-time algorithm
unlikely. The same problem arises when comparing networks
in terms of the trees they display~\cite[Section 6.14.3]{huson2010phylogenetic}.
Another group of proposals for metrics on phylogenetic networks
includes the $\mu$-distance~\cite{bai2021defining,cardona2024comparison}, as
well as the
\emph{nodal
distance} and \emph{triplet distance}~\cite{cardona2008metricsII}. 
One could loosely describe them as comparing the ``connectivity'' induced
on leaves by the networks topologies. While they are polynomial
to compute, they are only valid on sub-classes of networks, namely
\emph{orchard networks}~\cite{cardona2024comparison} and 
\emph{time-consistent tree-child networks}~\cite{cardona2008metricsII}.
A notable exception is the \emph{subnetwork distance}~\cite[Section 6.14.4]{huson2010phylogenetic}, defined as the symmetric difference of the
sets of \emph{rooted subnetworks}, which is valid on all networks,
and can be computed in polynomial-time for bounded-degree networks.

However, contrary to the Robinson-Fould metric on trees,
none of the measures mentioned above allow to outline a single
\emph{common sub-structure} in input networks (at least not directly). 
A natural way to obtain metrics valid on the entire space of phylogenetics is
to use \emph{operational definitions}, i.e.\,to define a distance as the
minimum number of a set of ``edition operations'' needed to transform one
network into another. Note that, as mentioned earlier, the original definition
of the Robinson-Foulds distance falls into this category.  On networks,
examples of such operations include \emph{nearest-neighbor interchange}
(NNI,~\cite{gambette2017rearrangement}), \emph{subtree prune and regraph}
(SPR,~\cite{bordewich2017lost}) or \emph{cherry-picking
operations}~\cite{landry2022defining} on orchard networks. While cherry-picking
operations have been successfully used to define and compute metrics on orchard
networks~\cite{landry2023fixed}, computing distances based on NNI+SPR
operations is NP-hard even for
trees~\cite{bordewich2005computational,dasgupta1998computing}. A remarkable
aspect of operational distances is also that, if some operations ``reduce'' the
network (as the edge contraction for Robinson-Foulds) and others ``expand'' it
(as the edge expansion for Robinson-Foulds) then a natural notion of ``maximum
common reduced network'' emerges as a way to compare networks.  We see it as
the best possible case of ``exhibiting common structure'' using a metric.
Interestingly, both cherry-picking operations~\cite{landry2023fixed} and
SPR~\cite{klawitter2020spaces} allow to define such a maximum common structure.
Another example of this philosophy is the computation of
\emph{maximum agreement sub-networks}~\cite{choy2005computing}, although
their definition is not operational.

In this work, we generalize the original definition of Robinson-Foulds, 
by defining and studying an operational metric for phylogenetic networks
based on edge contractions and expansions. We first prove
the connectivity of the space of networks having the same leaf sets under these operations. Then, we use them to
define two possible ways of comparing networks: (1) through the minimum number of 
contractions and expansions connecting two networks and (2) 
based on the size of a maximum common contraction. We show
that the former defines a metric, and the latter a \emph{semi-metric},
i.e.\,a dissimilarity measure verifying all properties of a distance except
the triangle inequality.
Given our purpose to outline common structures between networks, we focus
on the computation of a maximum common contraction (MCC) semi-metric. 
To do so, we first provide a characterization
of contracted networks in terms of a \emph{witness structure}, as well as 
a result regarding the \emph{diameter} of the maximum common contraction
semi-metric.
We also prove
its NP-hardness under different conditions: when restricted to bounded-degree
networks and a common contraction of size $\leq 3$ (number of non-leaf nodes), and when one of the networks
has only $5$ leaves, and the other has unbounded degree. However,
we also provide a polynomial-time algorithm for the case of weakly galled trees.

% The next sub-section discusses the position of our work relative to the graph
% algorithmics literature, in which the subject of graph contraction has been
% intensely studied.  Then, in Section~\ref{sec:foundations}, we formally define
% edge contractions and expansions, and the two ways they may provide metrics on
% phylogenetic networks. We also prove there the connectivity of the space of all
% phylogenetic networks under edge expansions and contractions.
% Section~\ref{sec:hardness} presents our hardness results, and
% Section~\ref{sec:algo_preliminary_work} and~\ref{sec:algo} our polynomial
% algorithm in the case of weakly-galled trees. Finally,
% Section~\ref{sec:conclusion} discusses some immediate implications of our
% hardness results from a parameterized complexity point of view, and identifies
% open problems. 
%All omitted proofs may be found in the Appendix, or in the full online pre-print.  \ml{TODO: remove last sentence if it becomes obsolete.}

\medskip\noindent\textbf{Related works outside of phylogenetics.}
Perhaps the closest related work in its philosophy is~\cite{zelinka1990contraction},
which defines a distance between (isomorphism classes of) undirected,
unlabeled graphs. However, no algorithmic results
are given. There does exist a 
rich line of work in algorithmic graph theory on contraction
problems in undirected, 
unlabeled graph~\cite{brouwer1987contractibility,levin2008computational,
kaminski2010contractions,belmonte2014parameterized,matouvsek1992complexity}. 
 The typical problem of interest in this literature is that
of deciding whether a graph $H$ is the contraction of a graph $G$
(either with $H$ fixed or not), not computing a common contraction between them. To the knowledge of the authors, no attention has
been given to this problem so far from
an algorithmic perspective. This is surprising in the light of the
fact that the \emph{maximum common subgraph}
problem, which is similar in philosophy (but completely different in its
definition), has been the subject of some 
work~\cite{kriege2018maximum,akutsu2020improved}.  We also highlight that we forbid contractions that create directed cycles, which differs from most previous work.  We also assume that leaves are uniquely labeled in the networks.  This is an important difference, as deciding
whether an unlabelled tree is the contraction of another unlabelled tree is NP-hard~\cite{matouvsek1992complexity}, whereas this is polynomial-time
solvable when leaves are labeled, as this is equivalent to computing the Robinson-Foulds distance. 

\section{Contraction-Expansion distance}
\label{sec:foundations}

\subsection{Preliminaries}

This article uses standard directed graph terminology (nodes, edges, in-neighbors, out-neighbors, acyclic graphs, $\dots$), as can for instance be found in
\cite{diestelbook}. 
In particular, we use the notion of DAG (directed acyclic graph) and topological
ordering. A \emph{topological ordering} of a DAG $\N$ is a permutation $\sigma =
(u_1, \ldots, u_n)$ of $V(\N)$ such that $(u_i, u_j) \in E(\N)$ implies $i < j$.
Hence, no backwards arc and thus backwards path is possible. Such an ordering
always exists in a DAG, and conversely, a directed graph allowing for a
topological ordering is a DAG.
We write $[n]$ for the set of integers
going from $1$ to $n$.  
A \emph{phylogenetic network} $\N=(V,E)$ 
is a directed acyclic graph (DAG) such that exactly one node (the \emph{root})
has no in-neighbors, and nodes with no out-neighbor (the \emph{leaves}) have in-degree
$1$. We may write \emph{network} for short. A non-leaf node is called an \emph{internal node}.  
We use $V(\N), E(\N), I(\N)$, and $L(\N)$ to denote, respectively, the set of nodes, edges, internal nodes, and leaves of $\N$. If there is an edge from $u$ to $v$, we may write either $(u, v)$ or $u \rightarrow v$.  A \emph{reticulation} is a node of
in-degree two or more.
In a network, an in-neighbor of a node is called a \emph{parent}, and an out-neighbor is a \emph{child}.  
The set of parents and children of a node $u \in V(\N)$ are respectively denoted by  $\innbr{\N}(u)$ and $\outnbr{\N}(u)$.
Note that a node may have multiple parents, and that we allow a root with a single child, as well as nodes with a single parent and a single child.
We say that $u$ \emph{reaches} a leaf $\ell$ if there exists a directed path
from $u$ to $\ell$. The set of leaves that $u$ reaches in $\N$ is denoted $D_\N(u)$ (or $D(u)$ for short).  If $\N$ is a tree, such a set $D(u)$ is sometimes called a \emph{clade} or \emph{cluster}.

%\nt{Finally,
%$\N$ is a \emph{binary network}
%if for each \nt{internal} node has total\nt{?} (in+out) degree $3$, except the root of out-degree $2$\nt{?} and leaves of in-degree $1$.}

To continue, developing metrics between networks requires in particular a notion
of \emph{equality} between them. To that end, we use the notion of
graph isomorphism, augmented with leaf preservation. Specifically, 
two networks $\N_1=(V_1,E_1)$ and $\N_2=(V_2,E_2)$
are said \emph{isomorphic}, denoted $\N_1 \sim \N_2$, if there exists a bijection 
$\phi:V_1\rightarrow V_2$ such that (a) $\forall u,v \in V_1$, $(u,v)\in E_1$ if and only
if $(\phi(u),\phi(v))\in E_2$ and (b) $\forall u\in L(\N_1), \phi(u) = u$.
Given this notion, we recall the definition of a \emph{metric} (or \emph{distance}) on phylogenetic networks. 
A function $d$, associating a real value to an arbitrary pair of input networks, is
a \emph{metric} if $\forall \N_1,\N_2$: (positivity) $d(\N_1,\N_2) \geq 0$; (identity) $d(\N_1,\N_2)=0$ if and only if $\N_1\sim \N_2$;
(symmetry) $d(\N_1,\N_2)=d(\N_2,\N_1)$; and (triangle inequality) $\forall \N_1,\N_2,\N_3$
$d(\N_1,\N_3)\leq d(\N_1,\N_2)+d(\N_2,\N_3)$.

\medskip
\begin{remark}[graph isomorphism]
%    Note that graph isomorphism (i.e.\,isomorphism on unlabelled graphs) contains phylogenetic
%    network isomorphism, as one can replace each leaf with a unique subgraph to force their identification.
    Note that graph isomorphism is polynomial-time solvable on \emph{bounded-degree} graphs~\cite{luks1982isomorphism},
    and therefore on bounded-degree networks as well\footnote{Isomorphism on networks is contained in the isomorphism problem on directed graphs, as
    each leaf may be replaced by a unique subgraph, implementing the same mapping
    constraint.}. 
\end{remark}

    Note that it is often the identity criteria which has limited
    %, in previous developments
    %of metrics on phylogenetic networks, 
    %has limited their applicability
    the applicability of metrics
    to sub-families of networks. For instance, \cite[Figure 6.28]{huson2010phylogenetic}
    gives an example of a pair of networks that are different, while
    their \emph{hardwired cluster}, \emph{softwired cluster} and
    \emph{tree containment} distances are $0$. A similar example is given
    in \cite{bai2021defining} for the $\mu$-distance.

\subsection{Contractions and expansions}

\begin{figure}
\centering
\includegraphics[width=\textwidth]{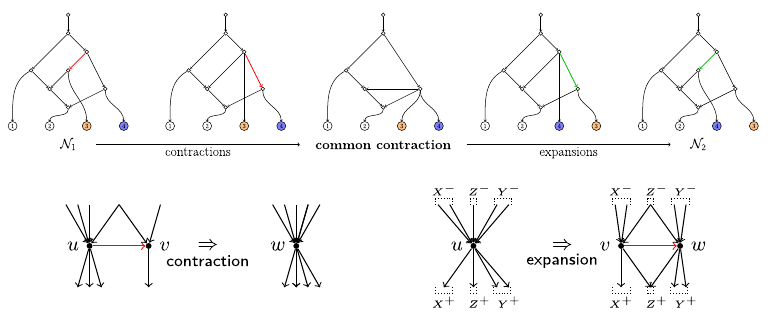}
    \caption{ (top) The transformation of a network into another
    through contractions, followed by expansions. 
    Midway, the intermediate network is a maximum common contraction, which can be achieved from $\N_2$ by reversing the expansions.
    (bottom) Illustration of edge contractions (left) and expansions (right). For expansions, the sets $X^-,Y^-,Z^-,X^+,Y^+,Z^+$
    specify how the neighbors of $u$ are distributed to $v$ and $w$.
    %They can be used to define two ways of comparing phylogenetic networks: (1) using the minimum number of contractions and expansions required to transform one network into another and (2) using the size of a maximum common
    %contraction. 
    Note that contractions may delete cycles and expansions create them.
    }
%    \bm{personally, I like this addition}}
    \label{fig:contraction_expansions}
\end{figure}

We now define edge contractions and expansions, as illustrated in
Figure~\ref{fig:contraction_expansions}, and the two ways we compare
networks using them.

\medskip
\begin{definition}[Contraction]
%    \ml{TODO: there should be some white space before the Def, Prop, Lem, etc.}
    Let $\N=(V,E)$ be a network and let $(u, v) \in E$. The contraction operation $c(u, v, w)$ 
    %\nt{[I am not convinced to include $w$ in the contraction notation because $(u,v)\in \N_1$ and $(w)\in \N_2$]} 
    on $\N$, where $w \notin V$, yields the
    directed graph $\N'=(V',E')$ such that:
    \begin{enumerate}
        \item 
        $V'= (V\setminus\{u,v\}) \cup\{w\}$; 
        \item 
        $\innbr{\N'}(w)=\innbr{\N}(u)\cup \left(\innbr{\N}(v)\setminus\{u\}\right)$; and 
        \item 
        $\outnbr{\N'}(w)=\left(\outnbr{\N}(u)\setminus\{v\}\right)\cup
    \outnbr{\N}(v)$.
    \end{enumerate}
    We denote by $\N / (u, v)$ the network obtained after applying the contraction $c(u, v, w)$ on $\N$.
    \label{def:contraction}
\end{definition}

A contraction $c(u,v,w)$ on a network $\N$ is said \emph{admissible} if 
$\N/(u,v)$ is also a network, on the same set of leaves. This means that 
$v$ is not a leaf, and that the contraction does not 
create a cycle. The following proposition characterizes the set of admissible contractions
in a network. 
This result can also be found in ~\cite[Lemma~3]{hellmuth2023clustering}.
%As this result can be found in~\cite[Lemma~3]{hellmuth2023clustering},
%we delay its proof to the Appendix.
\medskip
\begin{prop}
 A contraction $c(u,v, w)$ applied to $\N$ creates a cycle if and only if there exists a directed path from
$u$ to $v$ that does not use edge $u\rightarrow v$.
\label{prop:admissible}
\end{prop}
\begin{proof}
    If there is such a path $u = p_1\rightarrow \dots \rightarrow p_\ell = v$ in $\N$ (with $\ell \geq 3$), then contracting edge $u\rightarrow v$ into a newly
    introduced node $w$ (i.e.\,operation $c(u,v,w)$) creates the cycle
    $p_1=w\rightarrow\dots\rightarrow p_\ell =w$.
    If there is no such path, we show that there is a topological order
    of $\N$ in which $u,v$ are adjacent (in that order).  Within such an order,
    by inserting $w$ at the position of $u,v$ and removing them, we get
    a valid topological order for $\N'$, the graph obtained by applying $c(u,v,w)$ 
    to $\N$, which is therefore acyclic. 

    Let us therefore start from a topological
    order $\sigma$ of $\N$, and let us denote by $S$ the nodes of $\N$ ordered
    between $u$ and $v$ in $\N$. We show that $S$ may always be reduced
    in size by a slight
    re-shuffling of $\sigma$, showing that it may be empty. 

    To start with, if no node in $S$ is an out-neighbor of $u$, then $u$ may be moved to make it adjacent to $v$ while keeping the
    topological order valid. If this is not the case, then let $x$ be an out-neighbor of
    $u$ in $S$. By assumption (no path between
    $u$ and $v$ not using $u\rightarrow v$), there may
    not be a path between $x$ and $v$. Consider the largest node $y$ in $S$
    (according to the topological ordering) such that there is no path from $y$ to $v$ in $\N$.
    %for which there is no path
    %to $v$, and let us call it $y$. 
    Note that such a $y$ must exist since, in particular, $x$ is in $S$ and does not reach $v$.  
    Any node $z$ after $y$ in $S$ must have a path
    to $v$. There may not be an edge from $y$ to such a $z$, as it would yield
    a path from $y$ to $v$. Therefore, $y$ may safely be moved after $v$
    %any such $z$,
    %and $v$, 
    in the topological ordering. This reduces $S$ by $1$. 
    This process may be repeated until $S$ is 
    empty.
\end{proof}

A network $\M$ is said to be a \emph{contraction} of a network $\N$ if there exists
a series of admissible contractions yielding, when applied on $\N$,
a network isomorphic to $\M$.  Also, $\M$ is a \emph{common contraction} of two networks
$\N_1$ and $\N_2$ if it is both a contraction of $\N_1$ and $\N_2$.

As for expansions, they essentially transform a node into an edge. 
 The definition is slightly more involved, as we must specify how the in and out-neighbors are split. Note that expansions may simply
be seen as the inverse operation of contractions. 

\medskip
\begin{definition}[Expansion]
Let $\N=(V,E)$ be a network and let $u\in V$. The expansion operation $e(u, v, w, X^-,Y^-,Z^-,X^+,Y^+,Z^+)$, such that 
every node of $\innbr{\N}(u)$ occurs in exactly one of $X^-, Y^-, Z^-$ and every node of $\outnbr{\N}(u)$ occurs in exactly one of $X^+, Y^+, Z^+$, 
%$X^-,Y^-,Z^-$ (resp. $X^+,Y^+,Z^+$)
%is a partition of $\innbr{\N}(u)$ (resp. $\outnbr{\N}(u)$)
    yields the network $\N'=(V',E')$ such that: 
    \begin{enumerate}
        \item 
        $V'=V\setminus\{u\}\cup\{v,w\}$, where $v, w$ are two new nodes; 
        \item 
        $\innbr{\N'}(v)=X^-\cup Z^-$, 
    $\innbr{\N'}(w)=Y^-\cup Z^-\cup\{v\}$, $\outnbr{\N'}(v)=X^+\cup Z^+\cup\{w\}$ and $\outnbr{\N'}(w)=Y^+\cup Z^+$.
    \end{enumerate}
\label{def:expansion}
\end{definition}

The sets $X^-,Y^-,Z^-,X^+,Y^+,Z^+$ specify how the neighbors of the original
node $u$ are distributed among the newly created ones $v$ and $w$.
Nodes in $X^-/X^+$ are attributed to $v$ only, those in $Y^-/Y^+$ to $w$ only,
and those in $Z^+/Z^-$ to both. Note that, compared to the definition
of these operations in the case of trees~\cite{robinson1981comparison},
we need to specify more information as to how the neighbors
of the original node $u$ are ``split'' between $v$ and $w$. 
It is quite straightforward to see that, if an expansion replaces $u$ with $(v, w)$, then the contraction $c(v, w, u)$ reverses it.  Conversely, a contraction $c(u, v, w)$ that yields node $w$ can be reversed with the expansion that specifies the $X, Y, Z$ parameters as the appropriate sets of previous in and out-neighbors of $u$ and $v$.
%that didn't work, because eg $N^-(v)$ must be replaced by $N^-(v) \setminus N^-(w)$.  That got way too messy, so I replaced with a sentence.
%$e(u,v,w,N^-(v),N^-(v)\cap N^-(w),N^-(w),N^+(v),N^+(v)\cap N^+(w),N+(w))$ reverses
%a contraction of $v\rightarrow w$ into $u$.
As for contractions, we define \emph{admissible expansions} on a
network $\N$ on $L$ leaves as those that yield a phylogenetic
network on $L$ leaves after application (no creation of a second
root, of new leaves, or cycles).

Based on these definitions, we define two ways of comparing phylogenetic
networks.  The first one is the \emph{contraction-expansion distance}, defined
as the least number of contractions and expansions required to transform one
network into another.  
%Thanks to its operational definition, proving that it is
%a \emph{metric} on networks is quite straightforward. 
The second one is a
dissimilarity measure based on the \emph{maximum common contraction} of two
networks (or more precisely, the distance to a common contraction).  Although
%(as we shall see in the next section) 
we will see that
the latter does not verify the triangle
inequality, and therefore does not qualify as a ``metric'', we still make it
the main subject matter of this article, given the common structure it outlines between two networks.  
%Let us now state both definitions, starting
%with the \emph{contraction-expansion distance}.

\medskip
\begin{definition}[contraction-expansion distance]
    Given two networks 
     $\N_1$ and $\N_2$ over the same leaf sets,
    the contraction-expansion distance $\dce(\N_1,\N_2)$
    is the minimum 
    %\nt{\sout{possible}} 
    length of a sequence
    of admissible contractions and expansions transforming $\N_1$
    into a network $\N_2'$ such that $\N_2'\sim\N_2$.
\end{definition}

As for the dissimilarity measure based on maximum common contraction, we call it $\delta_{\mcc}$,
emphasizing that it is not a distance by not using $d$.

\medskip
\begin{definition}[MCC dissimilarity measure]
    Given two phylogenetic networks  $\N_1$ and $\N_2$ over the same leaf sets,
    the maximum common contraction (MCC) dissimilarity measure  $\delta_\mcc(\N_1,\N_2)$
    is defined as $\delta_\mcc(\N_1,\N_2)=|I(\N_1)|+|I(\N_2)|-2|I(\M)|$
    where $\M$ is a common contraction of $\N_1,\N_2$ of maximum size.
\end{definition}

\medskip
\begin{remark}
    As a common contraction also provides a sequence of contractions and expansions connecting
    both networks (by inverting one of the list of contractions into expansions), we have
    $\dce(\N_1,\N_2)\leq \deltamcc(\N_1,\N_2)$.
\end{remark}

Our main algorithmic problem of interest, formulated as a decision problem, is
the following: given a pair of networks $\N_1$ and $\N_2$ on the same leafset
and an integer $k$, is there a common contraction of size larger than $k$? 

\begin{mdframed}[nobreak=true]
    \textsc{Maximum Common Network Contraction (MCNC)}\\
    \textbf{Input:} Networks $\N_1,\N_2$, integer $k$\\
    \textbf{Question:} Is there a common contraction of $\N_1$ and $\N_2$
    with more than $k$ internal nodes?
\end{mdframed}

Note that it is equivalent to asking whether $\delta_\mcc(\N_1,\N_2)\leq k'$, with $k'=|I(\N_1)|+|I(\N_2)|-2k$.
%Before diving into the algorithmic study of \mcmc, 
We next establish important properties of
$\dce$ and $\delta_\mcc$. 
%Namely, we prove that $\dce$ is a metric, and provide examples
%showing that $\dce\neq \delta_\mcc$ and that $\delta_\mcc$ does not verify the triangle inequality.

\subsection{Properties of $\dce$ and $\delta_\mcc$}
\label{subsec:properties}

We first establish that admissible contractions and expansions
connect the set of all
phylogenetic networks on the same set of leaves. Within it,
the \emph{star network} denotes the network consisting of a single non-leaf
node to which all leaves are connected.

\medskip
\begin{prop}
Any phylogenetic network $\N$ may be contracted into the star network
by a series of admissible contractions.
    \label{prop:star}
\end{prop}
%
%\begin{proofsketch}
%    Let $r$ be the root of $\N$.  Because $\N$ is an acyclic digraph, $r$ has a child $u$ such that no other child of $r$ reaches $u$ (here, $u$ would be next to $r$ in a topological ordering).  Thus, there is no other path from $r$ to $u$, meaning that $r \rightarrow u$ is admissible by Proposition~\ref{prop:admissible}.  We can thus keep doing contractions from the root until a star is achieved.  
%\end{proofsketch}
%
\begin{proof}
We argue that the contraction of the edge between the root $r$ of $\N$ and its first non-leaf 
out-neighbor in any topological
ordering is always admissible. Let $\sigma$ be any
    topological ordering of $V(\N)$, 
    and let $u$ be the first non-leaf out-neighbor of the root $r$ that occurs
    in $\sigma$ (if no such out-neighbor exists, then $\N$ is already a star). If there exists a directed path between $r$ and $u$ that does
    not use edge $r\rightarrow u$, then any node on this path must be between
    $r$ and $u$ in $\sigma$, which goes against the definition of $u$.
    Therefore, by Proposition~\ref{prop:admissible}, $r\rightarrow u$ is
    admissible. It follows that we can always find a contraction of an edge incident to the root, and thus $\N$ can then be contracted into the star
    network by picking such an edge as long as there are non-root
    internal nodes.
\end{proof}

The corollary below uses Proposition~\ref{prop:star}
and the star network to show that one can always transform
a network into another. 
%Given two networks $\N_1$ and $\N_2$, we can therefore
%get two sequences of admissible contractions $C_1$ and $C_2$
%transforming each into the star network. Inverting $C_2$
%and appending it to $C_1$, we get the following corollary:
%Formally, two networks are identical
%when they are isomorphic. Let us therefore quickly describe how
%an isomorphism between
%two networks may be ``tracked'' upon applying analog contractions
%and expansions on both networks. To be more specific,
%given $\N_1,\N_2$ two isomorphic networks, $\phi$ an isomorphism
%between them and $c(u,v,w)$ a contraction operation on $\N_1$,
%the operation $c(\phi(u),\phi(v),w'$ is called
%the \emph{analog} of $c(u,v,w)$ in $\N_2$. It can easily
%be verified that applying $c(u,v,w)$ on $\N_1$ and its analog
%on $\N_2$ results in two isomorphic networks $\N_1',\N_2'$,
%and that an isomorphism between them is $\phi'$, equal
%to $\phi$ on all nodes $\notin\{u,v\}$ and such that
%$\phi(w)=w'$. Similar manipulations can be done
%to ``track'' an isomorphism through an expansion. In fact,
%one can even see contractions and expansions as being
%applied on \emph{isomorphism classes} instead of
%individual networks.
\medskip
\begin{corol}
Given any two networks $\N_1$ and $\N_2$ over the same leafset, there always exists a common contraction, and there always exists 
a series of admissible contractions and expansions that transforms $\N_1$ into
$\N_2'\sim \N_2$.  
\label{corollary:connectivity}
\end{corol}
\begin{proof}
    Through Proposition~\ref{prop:star},
    let $C_1$ and $C_2$ two sequences of contraction
    operations transforming $\N_1$ and $\N_2$
    into two isomorphic star networks on the same set of leaves as $\N_1$ and $\N_2$.  This shows that a common contraction exists.  Moreover, 
    by reversing $C_2$ into a sequence
    of expansions, appending it to $C_1$,
    and tracking the isomorphism $\phi$ along
    the resulting sequence of contractions and 
    expansions, we get a sequence of admissible operations
    transforming $\N_1$ into $\N_2'\sim \N_2$. 
\end{proof}

\begin{remark}
To finish, note that 
the length of the sequence 
described above is $|I(\N_1)|+|I(\N_2)|-2$, which is an upper bound on the distance.
\label{remark:max_dist}    
\end{remark}

We will see further down in this section that this value is in fact (under
a varied set of constraints) the \emph{diameter} of $\deltamcc$.

It can then be shown that $\dce$ is a metric. 
Due to its ``operational'' definition, it is relatively straightforward.  In particular, the triangle inequality is satisfied, since given three networks $\N_1, \N_2, \N_3$, we can always transform $\N_1$ into $\N_2$, then $\N_2$ into $\N_3$.  
\medskip
\begin{prop}
    $\dce$ is a metric on the set of phylogenetic networks
    with the same leaf sets.
\end{prop}
\ifcomplete
\begin{proof}
\emph{Positivity.}  This is trivially met since one cannot make a negative amount of operations.
\emph{Identity.} By definition, if $\N_1$ and $\N_2$
are isomorphic, $\dce(\N_1,\N_2)=0$. Conversely,
if $\N_1$ and $\N_2$ are not isomorphic, then at least
one operation is required to modify $\N_1$ into a network isomorphic to $\N_2$, and $\dce(\N_1,\N_2)>0$.\\
\emph{Symmetry.} Let $\Sigma$ be a sequence of admissible contractions
and expansions of minimum length transforming $\N_1$
into $\N_2'\sim\N_2$. By inverting $\Sigma$ into $\Sigma'$ and
tracking the isomorphism between $\N_2'$ and $\N_2$
along $\Sigma'$, we get a sequence of admissible
operations of length $|\Sigma|$ transforming $\N_2$ into a network $\N_1'$ isomorphic to $\N_1$.\\
\emph{Triangle inequality.} 
Let $\N_1, \N_2, \N_3$ be three networks on the same leafset.  Since any of these networks can be transformed into the other by Corollary~\ref{corollary:connectivity}, one can transform $\N_1$ into $\N_2$, then transform $\N_2$ into $\N_3$, which justifies $\dce(\N_1, \N_3) \leq \dce(\N_1, \N_2) + \dce(\N_2, \N_3)$.
\end{proof}
\fi

\begin{figure}
    \centering
    \includegraphics[width=\textwidth]{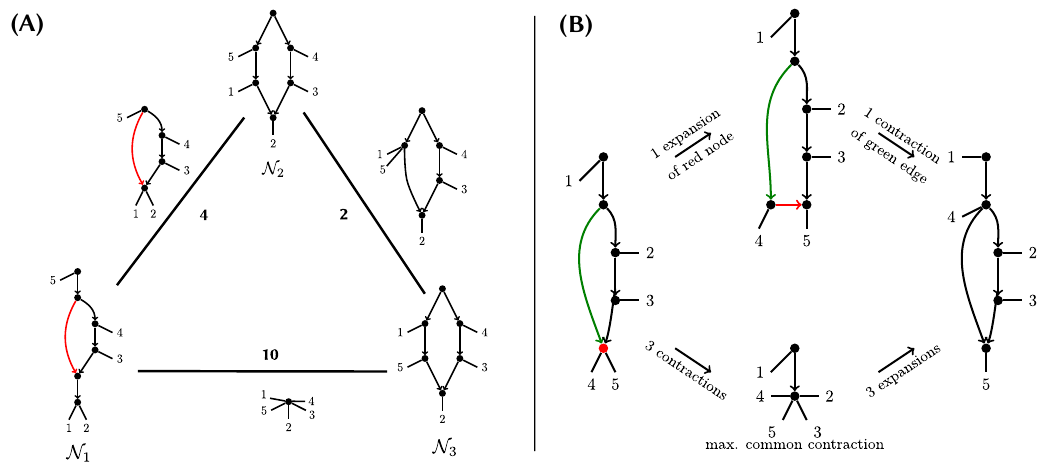}
    \caption{
    %\ml{TODO: I think you have a version of (A) with the networks displayed in a triangle, no?  The one you used for the talk.  If it's easy to integrate, I would use that figure.}
    (A) Examples of $3$ networks for which
    $\deltamcc(\N_1,\N_3)>\deltamcc(\N_1,\N_2) + \deltamcc(\N_2, \N_3)$.
    Indeed, as shown on the figure, there is a  common contraction
    of size $4$ between $\N_1$ and
    $\N_2$, and of size $5$ between $\N_2$ and $\N_3$. 
    However, the only possible common contraction between
    $\N_1$ and $\N_3$ is the star network.  This is due the
    edge highlighted in $\N_1$ not being admissible, enforcing the contraction of the whole cycle.
    Since all networks have $6$ internal
    nodes, we have $\deltamcc(\N_1,\N_3)=12-2=10$ whereas
    $\deltamcc(\N_1,\N_2)=12-8=4$ and $\deltamcc(\N_2,\N_3)=12-10=2$.
    (B) An example further highlighting the difference between $\dce$ and
    $\deltamcc$, as we have there two networks $\N_1,\N_2$ such that
    $\dce(\N_1,\N_2)=2$ (following the top path) whereas
    $\deltamcc(\N_1,\N_2)=6$ (following the bottom path). 
    }
    \label{fig:mcc_ce_diff}
\end{figure}

As for $\delta_\mcc$, it does verify identity ($\delta_\mcc(\N_1,\N_2)=0$ if and
only if their common contraction is themselves, i.e.\,they are isomorphic),
positivity, and
symmetry (by definition). However, Figure~\ref{fig:mcc_ce_diff}~(A) shows
an example in which $\delta_\mcc(\N_1,\N_3)=10$, while
$\delta_\mcc(\N_1,\N_2)+\delta(\N_2,\N_3)=4+2=6$, i.e.\,not verifying the
triangle inequality. 
Therefore, $\deltamcc$ is a \emph{semi-metric}.
We also highlight the difference between $\dce$
and $\deltamcc$ by showing on Figure~\ref{fig:mcc_ce_diff}~(B) an
example of two networks for which $\dce(\N_1,\N_2)=2$, whereas
$\deltamcc(\N_1,\N_2)=6$.  

\medskip
\begin{remark} 
If we allow contractions that create cycles, then we could show
that expansions and contractions can commute. This would imply that
contractions can always be carried out before expansions, and therefore
$\dce=\deltamcc$.  The fact that $\deltamcc$ is a distance on unlabelled,
undirected graphs with the same number of nodes has also been
shown~\cite{zelinka1990contraction}.
\end{remark}

\subsection{Witness structure to a contraction}

So far, we have only looked at contractions from an operational point of view, in which a network
is gradually changed into its contraction. For the purpose of easing formal proofs
in the following sections, we borrow from the undirected graph contraction literature (see for instance~\cite{belmonte2014parameterized}) the concept of \emph{witness structure}
of a contraction. 
%For two undirected graphs $G$ and $H$, it characterizes what sort of partition of the nodes of $G$ must exist in order for a contraction into $H$ to be possible.
The main idea is that, for a network $\N$ with contraction $\M$, a node $v$ of
$\M$ corresponds to a connected set of nodes in $\N$, that we call $W_v$. 
%\nt{ [The definition of $W_v$ is
%presented only after its initial usage or I don't find where.]} 
%\bm{[It was not really a ``usage", but more its first informal presentation. I added some more details.]}
The intuition is that the nodes of $W_v$ are
contracted together to become $v$.  Contractions impose $W_v$ to be
connected if we ignore edge directions (i.e.\,weakly connected), and the sets of nodes $\{W_v\}_{v\in I(\M)}$
must collectively partition $V(\N)$.
%\nt{I propose the following formalism to avoid
%contraction and, above all, to prevent any ambiguity regarding ‘. the sets of
%nodes $W_v$ must partition $V(\N)$}.

\medskip
\begin{definition}
Given two phylogenetic networks $\N$ and $\M$ on the same set of leaves, we
call an \emph{$\M$-witness structure in $\N$}
any partition $\W=\{W_u\mid u\in I(\M)\}$ of the internal nodes of $\N$ such that:
\begin{itemize}
    \item[(1)] $\forall u\in I(\M)$, $W_u$ is a non-empty weakly connected sub-graph of $\N$.
    \item[(2)] $\forall u,v\in I(\M)\times I(\M)$, $(u, v) \in E(\M)$ iff $\exists x\in W_u, y\in W_v$ such that $(x, y)\in E(\N)$.
    \item[(3)] $\forall u\in L(\M)$, the parent $p_u$ of $u$ in $\N$ must be in $W_{p'_u}$, where $p'_u$ is the parent of $u$ in $\M$.
\end{itemize}
\label{def:witness_structure}
\end{definition}

Note that as $I(\N)$ and $I(\M)$ are the sets of internal nodes, 
%\ml{[TODO: those sets also include leaves, so what did we mean by that?]}\bm{[This was a typo/mistake. I changed to I (and currently checking everything is consistent)]}, 
conditions (1) and (2) do not enforce an agreement on the placement
of the leaves between the contraction of $\N$ and $\M$. 
To obtain the equivalence between the presence of an $\M$-witness structure
and the contractibility of $\N$ into $\M$, condition (3) is
therefore needed.

In the case of undirected graphs, the equivalence between the presence of an
$\M$-witness structure and the contractibility into $\M$ is trivial. In our case
though, we need to guarantee that, although the connectivity of the sets in a
witness structure is only required in a weak sense (i.e.\,ignoring directions),
the presence of a $\M$-witness structure in $\N$ is still equivalent to
$\M$ being a contraction of $\N$, with all intermediary networks being acyclic.  This can be proved using similar ideas as in Proposition~\ref{prop:star}.

%\ml{[ML Note to ML: I skipped to section 3, must backcheck the following.]}
\medskip
\begin{prop}
    Let $\M$ and $\N$ two networks over the same set of leaves. Then $\M$ is a contraction
    of $\N$ if an only if there exists a $\M$-witness structure in $\N$.
    \label{prop:witness_structure}
\end{prop}
%\begin{proofsketch}
%    In the $(\Rightarrow)$ direction, if $v \in V(\M)$ is a node of the contraction, reversing the contractions ``re-expands'' $v$ to a witness set $W_v$, which must be weakly connected as each expansion adds an edge. 
%    In the $(\Leftarrow)$ direction, in each witness set $W_u$, the first edge in a topological ordering
%    is always admissible to contraction, and since it is connected, we may contract it to a single node. 
%    %Indeed, in the opposite case, there would
%    %be a directed cycle in $\M$, which is not possible (as it is a network).
%\end{proofsketch}
\begin{proof}
    \textbf{From contraction sequence to $\W$.}
    We work by induction on the length of the contraction sequence. For an empty
    contraction sequence from $\N$ to $\M$, $\N$ and $\M$ are isomorphic. 
    An $\M$-witness structure in $\N$ is given by $W_u=\{\phi(u)\}$ for all $u$
    internal node of $\M$, and $\phi$ isomorphism from $\M$ to $\N$.

    Let now $C=c_1,\dots, c_p$ be a contraction sequence from $\N$ to $\M$.
    Let us call $\M'$ the network obtained by applying the first $p-1$ contractions
    to $\N$. By the induction hypothesis, there is a $\M'$-witness structure $\W'$
    in $\N$. Let $c(x,y,z)$ be the $p$-th contraction. We remove $W_x$ and
    $W_y$ from $\W'$, and add to it a new set  $W_z=W_x\cup W_y$. We call
    the result $\W$ and argue it is a $\M$-witness structure in $\N$.
    Let us verify each point from the definition. $(1)$ is verified
    as $W_x$ and $W_y$ were both weakly connected, and are connected
    together by edge $x\rightarrow y$. $(2)$ is inherited from $\W'$ for any
    pair of nodes not involving $z$. For $u$ internal node of $\M$,
    $u\rightarrow z$ is an edge if and only if $u\rightarrow x$ or
    $u\rightarrow y$ were edges, which is equivalent to the existence of some edge
    from $W_u$ to $W_x$ or $W_u$ to $W_y$, in turn equivalent to the existence
    of an edge between $W_u$ and $W_z$. The case of an edge outgoing from $z$ is treated
    similarly. As for $(3)$, any leaf $u$ whose parent is $x$ or $y$ in $\M'$ now has $z$
    has a parent, and indeed, denoting by $p_\N(u)$ the parent of $u$ in $\N$, $p_{\N}(u)\in W_x $ or $p_{\N}(u) \in W_y$ implies $p_{\N}(u)\in W_z$.

    \textbf{From $\W$ to contraction sequence.} Given an $\M$-witness structure
    in $\N$ called $\W$, we prove that each set of $\W$ may be contracted into a single node.
    The result will then trivially be a network isomorphic to $\M$ (if $p_u$ is the node
    obtained from contracting $W_u$ then $\phi: p_u\rightarrow u$ is an isomorphism).
    We prove the result for each $u$ by induction on $|W_u|$.

    Consider therefore $W_u\in\W$, for $u$ internal node of $\M$,
    and the sub-graph $H_u$ induced by $W_u$ in $\N$. 
    We show that there is always an admissible
    contraction in $H_u$. Indeed, $H_u$ must be acyclic since a cycle in $H_u$ would
    be a cycle of $\N$. Consider the first node $x$ in a topological ordering of 
    $H_u$, and $y$ its first neighbor (in $H_u$) according to the same topological ordering.
    Note that unless $W_u$ is already a single node, such a $y$ must exist, as otherwise $x$
    has both no in-neighbor and no out-neighbor in $H_u$,  and thus $H_u$ is not weakly connected.
    By Proposition~\ref{prop:admissible}, the contraction of $x\rightarrow y$ is admissible
    if and only if there is no directed path from $x$ to $y$ not using edge $x\rightarrow y$.
    Such a path within $H_u$ is not possible, as then $y$ would not be the first
    neighbor of $x$ in a topological ordering. Outside of $H_u$, such a path is not possible
    either: let $u_1,\dots,u_t$ be, in order, the list of nodes in $\M$ whose corresponding set 
    ($W_{u_1},\dots,W_{u_t}$) is visited on such a path. By equivalence between the presence
    of an edge between $W_u$ and $W_v$ and the presence of $u\rightarrow v$ in $\M$,
    $u,u_1,\dots,u_t,u$ is a directed cycle of $\M$, which is not possible.
    Overall, there is always an admissible contraction in any set of a $\M$-witness
    structure. As the result is still a $\M$-witness structure, now in $\N/(x\rightarrow y)$, and with $|W_u|$ reduced by $1$, $W_u$ is contractible
    to a point by the induction hypothesis.
    We do the same for each $W_u$ to conclude the proof.
\end{proof}

\subsection{Diameter of $\deltamcc$}

The maximum value a distance method can produce when comparing phylogenetic
networks, called the diameter, is crucial to its practical use.
Indeed, it enables the normalization of distance values, 
allowing to quantify how two networks are similar relative to
a worst possible case.
%which is essential for comparing the results of different distance
%methods. This normalization is especially valuable when evaluating new methods
%against established ones. 
Specifically, given sets of networks $F_1,F_2$, we define the \emph{diameter over $F_1,F_2$}
for $\deltamcc$
as $$\text{diam}_{F_1,F_2} = \max_{\N_1,\N_2\in F_1\times F_2} \deltamcc(\N_1,\N_2)$$
A natural choice for $F_1,F_2$ could be the set of all networks 
over a given number of leaves. 
However, as we shall see, this yields a value of $+\infty$. To avoid it,
we look at the diameter obtained when restricting $F_1$ and $F_2$ to having
fixed (but different) numbers of internal nodes (equivalently, in binary networks, fixed numbers of reticulations).

In this setting, we show below that 
$\text{diam}_{F_1,F_2}$ is $|I(\N_1)|+|I(\N_2)|-2$, corresponding to the case where the common
contraction between both networks is the \emph{star network},
as defined in Section~\ref{subsec:properties}.
Note that this value matches the upper bound outlined in Remark~\ref{remark:max_dist}.
In other words, even when restricting $\N_1$ and $\N_2$ to having fixed numbers of nodes,
it is still possible to find a pair meeting this upper bound.

\medskip
\begin{prop}
Let $\ell, m, m'$ be integer numbers with $\ell>3$, $m$, $m'>1$. Then there are networks $\N,\N'$ with $l$ leaves such that $|I(\N)|=m, |I(\N')|=m'$, and $\delta_{MCC}(\N,\N')= m+m'-2$.
\label{prop:diameter}
\end{prop} 
%\bm{[BM: making it a normal proof, even if (right now) it means putting it in the appendix,
%we need to pass over every proof afterwards anyways to see which we integrate.`]}
\begin{proof}

 \begin{figure}
\centering
\includegraphics[width=\textwidth]{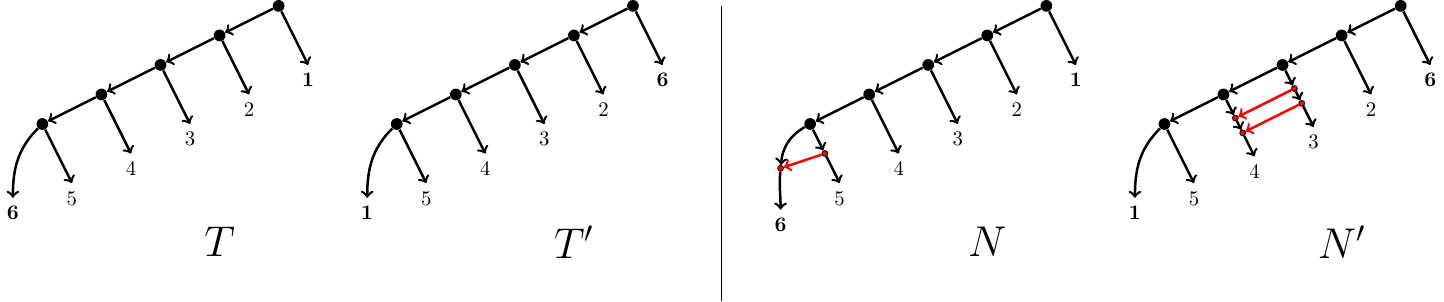}
\caption{(left) Two caterpillar trees over $7$ leaves whose common contraction is the star network. One way of seeing
it is that in $T$, leaf $1$ is a child of the root, and therefore this must be the case in a contraction of $T$ as well.
In $T'$, the only way to achieve this is to contract the parent of leaf $1$ into the root, making the star network
the only possible common contraction. (right) Reticulation edges have been added in $T,T'$ between different pairs of leaves
($P_N=\{(2,3)\}$ and $P_{N'}=\{(3,4)\}$ in the notations of the proof), in order to get $N,N'$ with respectively $7$ and $9$
internal nodes, while maintaining the property that the star network is the only possible common contraction.} 
\label{diam}
\end{figure}
Consider the caterpillar trees $T,T'$ in Figure~\ref{diam}~(left). Specifically, $T$ consists
of a directed path $P=p_1,\dots,p_{\ell-1}$ of length $\ell-1$, in which $\forall i\in [\ell-1]$, $p_i$ is the parent of 
leaf $i$, and $p_{\ell-1}$ is also the parent of leaf $\ell$. As for $T'$, it is built similarly (we call the path $P'$), except that leaves $1$ and $\ell$ are swapped. 
We argue that their maximum common
contraction $M$ is a star network. We will use the following Claim:
\\~\\
\textbf{Claim:} If a network $\M$ is a contraction of a network $\N$, with witness structure $\W=\{W_u\mid u\in\M\}$, 
then the root of $\N$ must be in the witness set $W_r$ associated to the root $r$ of $\M$.

\textbf{Proof of Claim:} First, note that $W_r$ is non-empty by definition of a
witness structure. Suppose the Claim is wrong, and consider the first node $x$
of $W_r$ in a topological ordering of $\N$. By assumption, $x$ is not the root.
It must therefore have an in-neighbor, which cannot be in $W_r$ (it would be
before $x$ in the topological ordering otherwise). By definition of a witness
structure, this means $r$ has an in-neighbor in $\M$, which cannot be the case, hence
a contradiction.
\\~\\
In the case of $T$ and $T'$ on Figure~\ref{diam}~(left), the Claim implies that
in any common contraction, leaf $1$ must be a child of the root (as it is the
case in $T$), as well as leaf $\ell$ (from $T'$). Therefore, in both
$T$ and $T'$, the parents of leaves $1$ and $\ell$ are both in the same witness
set (associated to the root of the contraction). As this witness set must be
connected, and there is only one path, consisting of all internal nodes,
connecting the parents of leaves $1$ and $\ell$ (both in $T$ and $T'$), the witness
set of the root contains all internal nodes. The maximum common contraction of
$T$ and $T'$ is therefore the star network.

%Indeed, in $T$, leaf $1$ is a child of the root.
%As $M$ is a contraction of $T$, and the root of 
%since in the common contraction the leaf $1$ should be a child of the root 
%%\bm{[This is true, but skips several steps. $1$ should be a child 
%%of the root because it is the child of the root in N1, and the witness set of the root
%%in a contraction must contain the root.]} 
%which requires
%contracting all the edges on the path between the root and $1$ in $T'$. 

We want to
construct $\N$ and $\N'$ by adding reticulated edges to the left-hand side tree $T$ and
the right-hand side tree $T'$ respectively, in a way that their maximum common
contraction remains the star network (see Figure~\ref{diam}~(right)). To be more
specific, we distinguish several
cases.

If $m=\ell-1$, then $\N$ is simply $T$. Likewise, if $m'=\ell-1$, then $\N'=T'$.
If $m < \ell-1$, then $\N$ is obtained from $T$ by contracting the last $\ell-1-m$ edges
of $P$. Likewise, if $m'<\ell-1$, $\N'$ is obtained from $T'$ by contracting the last $\ell-1-m$ edges
of $P'$. If $m>\ell-1$, and the difference is an even number, then $\N$ is constructed
from $T$ by taking the leaves $(\ell-1,\ell)$ (their parent is $p_{\ell-1}$), and
adding ``reticulation edges'' between them. Specifically, starting from $T$,
the edges $(p_{\ell-1},\ell-1)$ and $(p_{\ell-1},\ell)$ are removed. 
Instead, two paths $P_{\ell-1}$ and $P_{\ell}$ 
of length $(m-\ell)/2$, starting both at $p_{\ell-1}$, and ending at
$\ell-1$ and $\ell$, are introduced. Both of these paths have $(m-(\ell-1))/2$ internal
nodes. We add a (reticulation) edge from the $i$-th node of $P_{\ell-1}$ to the $i$-th
node of $P_l$, for $i\in [(m-(\ell-1))/2]$. This yields a network
with $\ell-1+2(m-(\ell-1))/2=m$ nodes. If $m'>\ell-1$, we build $\N'$ from $T'$ similarly,
except that we pick any pair $(u,v)$ distinct from $(\ell-1,\ell)$ (we may, as $\ell>3$), 
and the paths $P_u$ and $P_v$ (of length $(m'-(\ell-1))/2$)
that are introduced respectively start at the parent of $u$ and 
the parent of $v$. Finally, if $m>\ell-1$ and the difference is odd, we build $\N$
as in the even case, except that one of the edges of $P$ is contracted (any will do). If $m'>\ell-1$
and the difference is odd, we also build $\N'$ as in the even case, and contract
one edge of $P'$. In any case, we obtain networks $\N$ and $\N'$ on $\ell$ leaves with respectively
$m$ and $m'$ internal nodes.

We now argue that in all cases, the star network is the only possible
common contraction between $\N$ and $\N'$ (which implies $\deltamcc(\N,\N')=m+m'-2$).
To start with, note that in $\N'$, in all cases ($m'<\ell-1$ or not) the parent
of $\ell$ is the root.

Let $\M$ be a common contraction of $\N$ et $\N'$ and $\W$
the witness structure of the contractibility of $\M$ into $\N$.
The witness set associated to the root of $\M$ in $\N$ is denoted $W$.
In $\N$, both the root (by the Claim) and the parent of $\ell$ must be in $W$.

If $m\geq \ell-1$, then $\N$ is a contraction of $T$, and the set of all internal nodes
of $\N$ constitutes the only directed path starting at the root and finishing at the parent
of $\ell$. All of it must be included in $W$, yielding the star network as the only possible
contraction.

If $m>\ell-1$, then let us call $x_\ell$ the parent of $\ell$ in $\N$ (it is the penultimate
node of $P_\ell$). Note that any node of $P_\ell$ or $P_{\ell-1}$ is on a directed path
from the root of $\N$ to $x_\ell$. Indeed, consider the $i$-th nodes on the paths 
$P_{\ell-1}$ and $P_\ell$: 
by taking the (potentially contracted on one edge) path
to $p_{\ell-1}$ (starting point of $P_{\ell}$ and $P_{\ell-1}$) then going to 
the $i$-th node of $P_{\ell-1}$, taking the edge to the $i$-th node of $P_{\ell}$, and going
to $x_\ell$, we have a directed path visiting them both.

To conclude, we argue that \emph{any directed path} from the root of $\N$ to $x_\ell$ 
must lie entirely in $W$. Indeed, if it is not included in $W$,
then some vertices in these paths are in witness sets
associated to other vertices of $\M$, which yields
a directed cycle starting and ending at the root in $\M$, hence a contradiction.
As any node on the path from the root of $\N$ to $p_{\ell-1}$ is on every path from 
the root to $x_\ell$, we have that all internal nodes of $\N$ are in $W$, and
the only possible common contraction between $\N$ and $\N'$ is the star network.

\end{proof}

%\begin{figure}
%\centering
%\includegraphics[width=10cm, height=8 cm]{example.png}
%\caption{The network $\N$ with $4$ internal nodes and the network $\N'$ with with $7$ internal nodes are constructed by contracting and adding  reticulated edges to $T,T'$ respectively. } \label{examp}
%\end{figure}

%We have demonstrated that establishing an upper bound for $\delta_{MCC}$ requires fixing both the number of leaves and internal nodes.
A consequence of Proposition~\ref{prop:diameter} is that if $F_1,F_2$ are both simply
defined as the set of all networks over a given number of leaves, $\text{diam}_{F_1,F_2}$ is $+\infty$ 
(indeed, if it was equal to a finite value $M$, then we could for instance choose $m=M$ and $m'=2$ to exceed it).
Specifying the number of internal nodes is therefore required to get a finite value.

In addition, note that in this proof, each reticulated edge that is introduced adds exactly
one reticulation to a network. Therefore, the construction also allows to specify the
number of reticulations of each networks while ensuring their common contraction is 
the star network. The reticulation edges could also be added between several disjoint pairs
of leaves in each network, allowing to specify the \emph{level} of both networks,
as well as the number of reticulations up to a limit of $\text{level}\times\text{\# of leaves/2}$ (as with such a construction, one may require 2 leaves per biconnected component).
In all of these cases, having two networks whose common contraction is a star (and therefore
$\deltamcc(\N_1,\N_2)=|I(\N_1)|+|I(\N_2)|-2$) is possible, showing the generality
of this value as a diameter.

\section{NP-hardness of finding maximum common contractions}
\label{sec:hardness}

In this section, we prove that the \mcmclong problem is NP-hard,
through reductions from the {\sc Set Splitting} problem (SP12 in
\cite{garey1979computers}).
%, also
%known as {\sc Hypergraph 2-Colorability} in the literature.
%Its definition is the following:
\medskip
\begin{mdframed}[nobreak]
    {\sc Set Splitting}\\
    \textbf{Input:} A set $X$, a family $\mathcal{A}$ of subsets of $X$\\
    \textbf{Question:} Is there a bipartition $X_1,X_2$ of $X$ such that
    $\forall S\in\mathcal{A}$, $S\cap X_1\neq \emptyset$ and $S\cap X_2\neq \emptyset$ ?
\end{mdframed}
\medskip
If $(X,\A)$ is a yes-instance, we say it is \emph{splittable}.
We then call $X_1,X_2$ the \emph{split} of $(X,\A)$.
Note that \setsplit may be seen as {\sc Positive Not-All-Equal-SAT},
i.e.\,CNF satisfiability with the additional requirements that
(1) negations are completely absent (``positive'', also sometimes called ``monotone'' in the literature) and (2) at least one 
variable is set to false in each clause. 
%The reduction consists
%in seeing $X$ as the set of variables, $\A$ as the clauses, and $X_1/X_2$
%as the variables assigned to true/false.  

We provide two reductions from \setsplit to \mcmclong. 
The first one has three interesting features. First,
it shows that \mcmclong remains NP-hard when $k$ (the size of the common
contraction) is equal to $3$. Second, 
noting that {\sc
Positive NAE-SAT} remains NP-hard with $3$ variables per-clauses ({\sc
Positive-NAE-3-SAT}) and at most $4$ occurrences of each variable in all
clauses~\cite{darmann2020simple}, it shows that \mcmc is NP-hard
on phylogenetic networks of bounded degree. Third, when not putting
any constraint on the 
number of occurrences of each element, it allows importing lower bounds
for {\sc Positive-NAE-3-SAT} conditional to ETH (Exponential-Time Hypothesis)
~\cite[Proposition 5.1]{antony2024switching}.
As for the second one, it shows that \mcmc remains hard even when one of the
network is a path on $4$ nodes, each parent to one leaf. Not only 
is this simple network a tree, but it also has a constant number of leaves.

\subsection{Hardness with common contraction of size $k=3$}

\begin{figure}
    \centering
    \includegraphics[width=\textwidth]{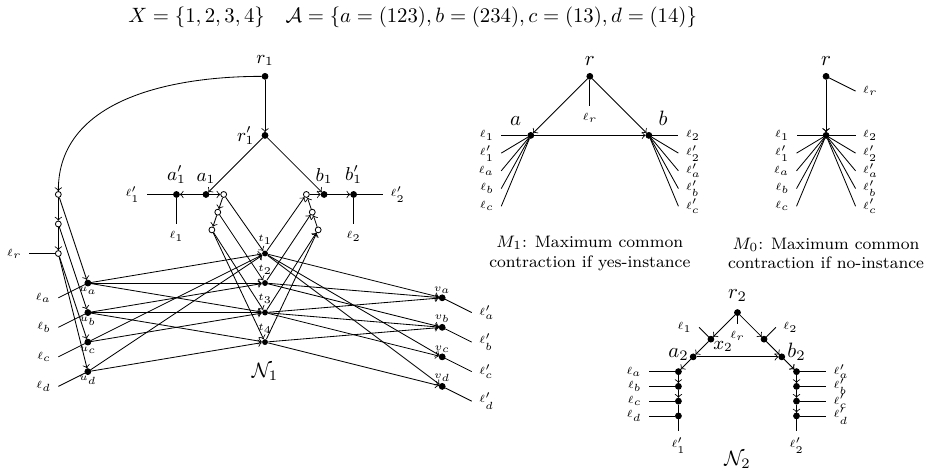}
    \caption{Illustration of our first reduction from \setsplit to
    \mcmclong on an example. 
    %\ml{[TODO: To be consistent, $\N_2$ should have arrows everywhere (except leaves)]}
    %\bm{[If I understood what you meant, done.]}}
    }
    \label{fig:np-hardness}
\end{figure}

Given an instance $(X,\A)$ of \setsplit, we build two binary phylogenetic
networks in the following way (see Figure~\ref{fig:np-hardness} for an
illustration on an example). First, we introduce two leaves $\ell_S$ and
$\ell_S'$ for each set $S\in\A$. We also add to this leaf-set two couples of leaves
$\{\ell_1,\ell_1'\}$ and $\{\ell_2,\ell_2'\}$, and a single leaf
$\ell_r$, whose positions in $\N_1,\N_2$
will allow us to enforce specific contractions. 

As for the networks, let us start with $\N_1$. Its purpose is to encode
the relationships
between the elements of $X$ and the sets in $\A$.
Its internal node set
is composed of one node $t_x$ for each $x\in X$, two nodes $u_S$
and $v_S$ for each $S\in\A$, a root $r_1$, and other nodes named
$r_1',a_1,a_1',b_1,b_1'$. We write $U=\{u_S\}_{S\in\A}$ and $V=\{v_S\}_{S\in\A}$.
For all $S\in \A$, the
parents of $\ell_S$ and $\ell_S'$
are $u_S$ and $v_S$, respectively. $\ell_1$ and $\ell_1'$ share a parent $a_1'$,
whose own parent is $a_1$ (see
Figure~\ref{fig:np-hardness}). The same connectivity pattern is set between
$\ell_2,\ell_2'$, $b_1'$ and $b_1$. 
The root has a child $r_1'$ which is the parent of $a_1$ and $b_1$, while
the other child is the first of a series of intermediary nodes (white circles on the left on Figure~\ref{fig:np-hardness}), called $W^O[r_1]$,
ensuring that $r_1\rightsquigarrow u_S$, $\forall S\in \A$.
Denoting $\A$ as $\{S_1,\dots,S_m\}$, $W^O[r_1]$ consists
of a path $p_1\rightarrow p_2 \rightarrow \ldots \rightarrow p_{m-1}$ of $m-1$ nodes,
%\ml{[this is edge notation, not path notation, maybe write eg $p_1 \rightarrow p_2 \rightarrow \ldots \rightarrow p_{m-1}$?]}, 
%<<<<<<< HEAD
%such that $\forall i\in [m-1]$, $p_i\rightarrow u_{S_i}$, $r_1\rightarrow p_1$,
%and $p_{m-1}$ is connected to $S_{m}$ in addition to $S_{m-1}$. 
%In addition, $p_{m-1}$ is the parent of $\ell_r$.
%Likewise,  
%$\forall x\in X$, we implement a directed path between $a_1$ and $t_x$, and
%between $t_x$ and $b_1$, using intermediary nodes called 
%$W^{O}[a_1]$ and $W^I[b_1]$, also arranged in a path of size $|X|-1$,
%with each node connected to one $t_x$ (except the last one, connected to $2$). 
%Their purpose is to implement directed
%=======
such that for $i\in [m-1]$, $p_i\rightarrow u_{S_i}$ is an edge.  
 Also add edges $r_1\rightarrow p_1$,
and $p_{m-1} \rightarrow S_m$.
In addition, we add a leaf $\ell_r$, such that $p_{m-1}$ is the parent of $\ell_r$.
%is connected to $S_{m}$ in addition to $S_{m-1}$. 
%Likewise,  
%$\forall x\in X$, we implement a directed path between $a_1$ and $t_x$, and
%between $t_x$ and $b_1$, using intermediary nodes called 
%$W^{O}[a_1]$ and $W^I[b_1]$, also arranged in a path of size $|X|-1$,
%with each node connected to one $t_x$ (except the last one, connected to $2$). 
%\ml{[TODO: I had a difficult time reconstructing the network with the previous description, so I felt the need to add all details, please backcheck.]}
In a similar manner, denoting $X = \{1, \ldots, n\}$, we add a set of nodes
$W^O[a_1]$ consisting of a path $q_1 \rightarrow \ldots \rightarrow q_{n-1}$,
such that $q_i \rightarrow t_i$ is an edge for $i \in [n-1]$, and add edges
$a_1 \rightarrow q_1, q_{n-1} \rightarrow t_n$.  Likewise, we add another set of nodes
$W^I[b_1]$ consisting of a path $q'_{n-1} \rightarrow q'_{n-2} \rightarrow
\ldots \rightarrow q'_1$ such that $t_i \rightarrow q'_i$ is an edge for $i \in
[n-1]$.  Finally, we add edges $t_n \rightarrow q'_{n-1}$ and $q_1 \rightarrow b_1$.
The purpose of these paths is to implement directed
paths with bounded degree nodes.
Finally and most importantly, we add an edge from $u_S$ to $t_x$, and
from $t_x$ to $v_S$, if and only if $x\in S$.

As for $\N_2$, it is a much simpler network. Its main feature
is the presence of a ``separation'' 
between $\{\ell_S\mid S\in\A\}$ and $\{\ell_S'\mid S\in \A\}$,
emulating the purpose of ``splitting'' each element of $\A$.
Similarly to $\N_1$, it contains nodes $r_2$ (root) and $a_2,b_2$.
The root $r_2$ is the parent of $\ell_r$, and two non-leaf children,  
which are themselves the parents
of $\ell_1$ and $a_2$, and $\ell_2$ and $b_2$, respectively. $a_2$
and $b_2$ are connected by an edge $a_2\rightarrow b_2$. Finally, $a_2$ $b_2$
are the parents of, respectively, a path of nodes with children 
$\{\ell_S\mid S\in \A\}\cup\{\ell_1,\ell_1'\}$
and $\{\ell_S'\mid S\in\A\}\cup\{\ell_2,\ell_2'\}$ (see Figure).

We prove below that $\N_1$ and $\N_2$ have a common contraction of size $\geq 3$ 
(more specifically, the network $M_1$ depicted on Figure~\ref{fig:np-hardness}) if
and only if $(X,\A)$ is splittable. 

\medskip
\begin{thm}
    \mcmclong is NP-hard, even when restricted to $k=3$
\end{thm}
%
%\begin{proofsketch}
%    Because $\ell_1,\ell_1'$ and $\ell_2,\ell_2'$ share a parent in $\N_1$,
%    it follows that $\N_2$ must be contracted to $M_1$ to reach any common contraction with $\N_1$.
%    The question is then whether $M_1$ is a possible contraction.
%    We consider therefore the existence conditions of a $M_1$-witness structure.
%    The essence of the proof is that to reach this witness structure, the $u_S$ vertices must be in the same witness $W_a$ set as $a_1$ and $a_1'$.  That witness set must form a weakly connected subgraph, and the $t_x$ must be used to connect each $u_S$ with $a_1, a_1'$ (as $r_1, r_1'$ must belong to another witness set).  A similar situation holds for the $v_S$ vertices and $b_1, b_1'$, which must be in the same witness set $W_b$.  Thus, $t_x$ must be split between $W_a$
%    and $W_b$, and this split may either yield a split for $(X,\A)$, or
%    be inferred from an existing split of $(X,\A)$.
%\end{proofsketch}
%
\begin{proof}
%\textbf{Details of $\N_1$ construction.}
%As touched upon before, we aim at implementing the necessary directed paths in
%$\N_1$
%with bounded-degree nodes. We do so by introducing intermediary nodes
%(denoted as white circles on Figure~\ref{fig:np-hardness}). 
%We adopt for them the following notations: the intermediary nodes
%introduced to allow the presence of directed paths starting at $u$ (resp. going to $u$)
%are denoted $W^O[u]$ (resp. $W^I[u])$. For instance, 
%for the implementation of a path $r_1\rightsquigarrow u_S$ $\forall S\in \A$,
%we first pick an arbitrary order $\sigma$ on $\{u_S\mid S\in\A\}$, and introduce
%a path $W^O[r_1]$ of intermediary nodes of size $|\A|-1$.
%Let us write $w_{r_1,i}$ the $i-th$ element along that path.
%We first introduce the edge $r_1\rightarrow w_{r_1,i}$. Then, $\forall i\in[|\A|-1]$,
%we introduce an edge from $w_{r_1,i}$ to the $i-th$ element of $\{u_S\mid S\in \A\}$ according
%to $\sigma$. Finally, the last element ($w_{|\A|-1,r_1}$) of $W^O[r_1]$ is connected to the
%last two elements of $\{u_S\mid S\in \A\}$ according to $\sigma$.
%A similar construction, depending on an arbitrary order $\sigma$ on $X$, implements the directed paths
%from $a_1$ to $t_x$ $\forall x\in X$ and from $t_x$ to $b_1$. 
%The sets of directed nodes are denoted, respectively, $W^{O}[a_1]$ and
%$W^I[b_1]$. 
    \textbf{Necessary contractions in $\N_2$.}
    We start the proof by showing that a common contraction of $\N_1$ and $\N_2$
    is at most the network depicted as $M_1$ on Figure~\ref{fig:np-hardness}.
    Indeed, note that $\ell_1$ and $\ell_1'$ share a parent in $\N_1$, whereas
    this is not the case in $\N_2$. Therefore, to get to any common contraction of $\N_1$
    and $\N_2$, one of the paths between $p(\ell_1)$ and $p(\ell_1')$ in $\N_2$ must be
    contracted to a single node. There are two such paths: one going through $b_2$ and $r_2$,
    and one going through the common neighbor of $a_2$ and $p(\ell_1)$, which we
    call $x_2$.  Both use the path from $a_2$ to $p(\ell_1')$, which
    must be contracted to a point. In addition, we argue that
    edge $p(\ell_1)\rightarrow x_2$ must be contracted. Indeed, if it is not,
    then the path from $x_2$ to $\ell_1$ going through $b_2$ an $r$
    must be contracted, but this would create a cycle (in the form of a self-loop). The same is true of leaves
    $\ell_2$ and $\ell_2'$.

    The result of the application of these necessary contractions on $\N_2$ 
    is no other than the network $M_1$ depicted on Figure~\ref{fig:np-hardness}.
    It is composed of three internal nodes $r,a,b$, three edges $r\rightarrow a$,
    $r\rightarrow b$ and $a\rightarrow b$, the set of leaves 
    $\{\ell_S\mid S\in\A\}\cup \{\ell_1,\ell_2\}$ 
    (resp. $\{\ell_S'\mid S\in\A\}\cup \{\ell_1',\ell_2'\}$) are connected to $a$ (resp. $b$),
    and leaf $\ell_r$ connected to $r$.
    A common contraction of $\N_1$ and $\N_2$ is therefore either $M_1$ or a contraction
    of $M_1$.

    \textbf{NP-hardness proof.}
    We use the fact that $M_1$ is a contraction of $\N_1$ if and only if
    there is a $M_1$-witness-structure in $\N_1$
    (Proposition~\ref{prop:witness_structure}). Recall that this is the case if
    one can partition the internal nodes of $\N_1$ into three sets $V_r,V_a,V_b$
    such that: (1) there exists edges going from $V_r$ to $V_a$, $V_r$ to $V_b$,
    and $V_a$ to $V_b$, (2) all other edges are internal to $V_a,V_b,V_r$,
    (3) leaves $\{\ell_S\mid S\in\A\}\cup \{\ell_1,\ell_2\}$ are connected to 
    $V_a$, leaves $\{\ell_S'\mid S\in\A\}\cup \{\ell_1',\ell_2'\}$ to $V_b$ and $\ell_r$ to $V_r$, and (4)
    $V_r,V_a,V_b$ each induce weakly-connected graphs in $\N_1$.
    We now prove that $M_1$ is a contraction of $\N_1$ if and only if $(X,\A)$ is splittable.

    \textbf{$(X,\A)$ splittable $\Rightarrow$ $M_1$ common contraction.}
    Let us first suppose that $(X,\A)$ is splittable, and let $X_1,X_2$ be a
    bipartition of $X$ such that $\forall S\in \A$, $S\cap X_1\neq \emptyset$
    and $S\cap X_2\neq\emptyset$. We first contract $W^O[r_1]$ into $r_1$, $W^O[a_1]$ into $a_1$ and $W^I[b_1]$ into $b_1$. 
    Now, $\forall x\in X$, there is an edge $a_1\rightarrow t_x$
    and an edge $t_x\rightarrow b_1$, as well as edges $u_S\rightarrow t_x$ and $t_x\rightarrow v_S$
    if and only if $x\in S$. We now contract $t_x$ into $a_1$ $\forall x\in
    X_1$, and $t_x$ into $b_1$ $\forall x\in X_2$. Due to the fact that $\forall
    S\in \A$,
    $X_1\cap S\neq \emptyset$, each $u_S$ is now connected to $a_1$ (likewise,
    each $v_S$ is connected to $b_1$). We contract these edges, as well as
    $r_1\rightarrow r_1'$, $a_1'\rightarrow a_1$ and $b_1\rightarrow b_1'$, to
    finish the contractions and obtain $M_1$.  
    %\ml{[TODO: if we have time, can
    %we argue that each contraction mentioned is admissible?]}

    \textbf{$M_1$ common contraction $\Rightarrow$ $(X,\A)$ splittable.}
    Conversely, let us suppose that $\N_1$ is contractible to $M_1$. There must
    therefore exist $V_a,V_b,V_r$ verifying properties (1),(2),(3),(4) stated above.
    To start with, because of the presence of the leaves $\ell_S$, we must have
    $u_S\in V_a$ (likewise, $v_S\in V_b$) $\forall S\in \A$. Also, $a_1'\in V_a$
    and $b_1'\in V_b$. As $V_a$ is
    connected, and $a_1$ is on all paths from $a_1'$ to any $u_S$, we also have
    $a_1\in V_a$ (and $b_1\in V_b$). This implies that
    each $t_x$ must be in $V_a$ or $V_b$, but not $V_r$,
    as all paths from $t_x$ to $r_1$ (which must be in $V_r$) are cut
    by nodes in $V_a$ or $V_b$. Finally, note that all of $W^O[r_1]$
    must be included in $V_r$, as $\ell_r$ must have its parent in $V_r$,
    and all paths from $p_{m-1}$ (the last node of $W^O[r_1]$ and parent of $\ell_r$) to $r_1$
    not using $W^O[r_1]$ are cut by vertices that must be in $V_a$ (namely, $\{u_S\}_{S\in\A}$).
    The only way to connect $p_{m-1}$ to $r_1$ is through $W^O[r_1]$ (which is a single path).

    Let us now define $X_1=\{x\in X\mid t_x\in V_a\}$ and $X_2=\{x\in X\mid
    t_x\in  V_b\}$. For all $S\in \A$, as $u_S\in V_a$, and all paths from $u_S$
    to $a_1$, excluding those that use $W^O[r_1]\cup\{r_1\} \subseteq V_r$, go
    through some $t_x$ for $x\in S$, we must have some $x\in S\cap
    X_1$ to ensure the connectivity of $V_a$. Likewise, since $v_S\in V_b$,
    there must be some $x\in S\cap X_2$ to ensure the connectivity of $V_b$.
    $X_1,X_2$ is therefore a valid split of $(X,\A)$.

%    \textbf{Membership in NP} 
%    \todobm{Use the ``witness structure'' defined in
%    section 2 as certificate.}
    
\end{proof}

Restricting the instances of \setsplit to having sets of size $3$ 
and exactly $4$ total occurrences of an element $x$ in $\A$ (a variant
that it still NP-hard~\cite{darmann2020simple}), we
get the following.

\begin{thm}
    \mcmclong is NP-hard, even with $k=3$ and networks of degree $\leq 10$.
\end{thm}
\begin{proof}
    We use the same reduction as above. Note that in the network $\N_1$
    constructed therein, all nodes except $\{t_x\}_{x\in X}$ and
    $\{u_S,v_S\}_{S\in \A}$ have (in+out) degree at most $4$ (achieved by
    the parent of $\ell_r$). 
    Each $t_x$ has degree $2$ times its number of occurrences in sets of $\A$,
    to which we add $2$ to account for the edge from $W^O[a_1]$ and to $W^I[b_1]$.
    With the restriction of \setsplit under which we work, this yields a degree
    of $10$. As for each $u_S$, their degree is exactly $5$ ($1$ edge from $W^O[r_1]$, $1$ to $\ell_S$, and $3$ to $t_x$ nodes). Likewise,
    each $v_S$ has degree $4$. From \cite{darmann2020simple}, we get the result.
\end{proof}

The complexity remains open for binary networks (i.e.\,with in+out degree=3 for every node).
Finally, we can argue that the number of nodes and edges of $\N_1$ and $\N_2$ produced by our reduction is linear with respect to $|X| + |\A|$.
The correspondence 
with {\sc Positive-NAE-SAT} allows to import ETH-based lower bounds~\cite[Proposition 5.1]{antony2024switching}:
\medskip
\begin{thm}
    Assuming the Exponential Time Hypothesis, \mcmclong cannot be solved in time $2^{o(|V(\N_1)|+|E(\N_1)|)}$.
    \label{thm:eth}
\end{thm}
\begin{proof}
    Proposition 5.1 in \cite{antony2024switching} states that, $\forall k\geq 3$
    and assuming ETH, {\sc Positive-NAE-$k$-SAT} cannot be solved in time
    $2^{o(n+m)}$
    (they use the term ``{\sc Monotone}'' as we use ``{\sc positive}'' in this
    article, i.e.\,signifying the complete absence of negations).

    We recall that the equivalence between a \setsplit instance $(X,\A)$
    and a {\sc Positive-NAE-$k$-SAT} instance
    simply consists in interpreting each element of $x$ as a variable
    and each set in $\A$ as a clause. Therefore $n=|X|$ and $m=|\A|$,
    and given an instance
    of {\sc Positive-NAE-$k$-SAT}, our reduction constructs a pair of network
    $\N_1,\N_2$, in which $|U|=|V|=m$
    and $|\{t_x\}_{x}|=n$. To these we need to add $2m+5$ leaves, $2(n-1)$
    nodes in $W^{O}[a_1]$ and $W^I[b_1]$, and $m-1$ nodes in $W^O[r_1]$.
    The remaining nodes constitute a set constant size. Thus $|V(\N_1)|=O(n+m)$.
    Likewise, there are $\leq 2km$ edges between $U,V$ and $\{t_x\}_{x\in X}$,
    $\leq m$ edges between $W^O[r_1]$ and $U$, and betwen $U,V$ and the leaves.
    To finish, there are $\leq n$ edges between $W^O[a_1],W^I[b_1]$ 
    and $\{t_x\}_{x\in X}$, the remaining set of edges being of constant size.

    Overall $|V(\N_1)|+|E(\N_1)|=O(n+m)$. Therefore if we had 
    an algorithm solving \mcmclong in time $2^{o(|V(\N_1)|+|E(\N_1)|)}$, it could be
    used to solve {\sc Positive-NAE-$k$-SAT} in time $2^{o(n+m)}$,
    which is not possible under ETH.
\end{proof}

\subsection{Hardness with $\N_2$ a path with $5$ leaves}

%\begin{toappendix}
\begin{figure}
    \centering
    \includegraphics[width=.8\textwidth]{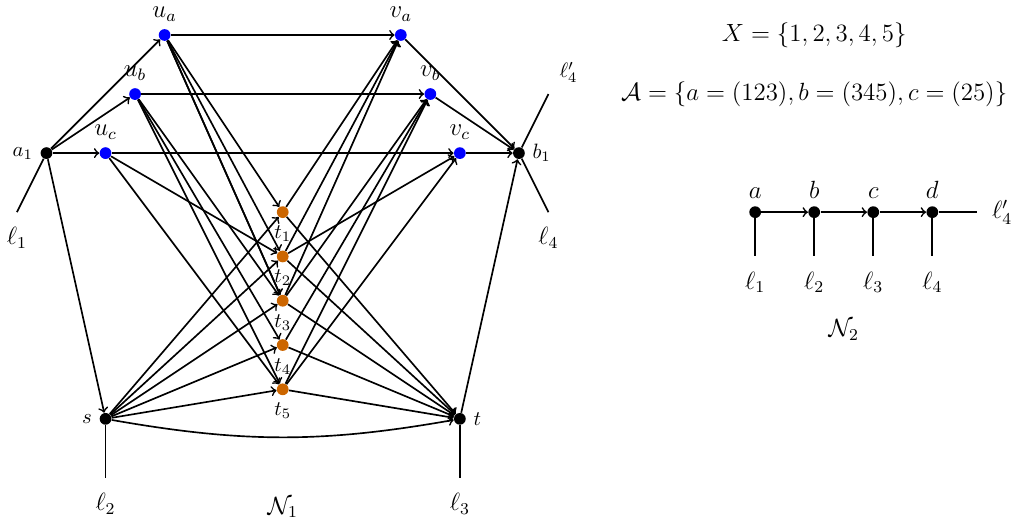}
    \caption{Illustration of our second reduction, also from 
    \setsplit. The blue nodes represent the sets of $\A$, and
    the orange nodes the elements of $X$. We use
    this construction in Theorem~\ref{thm:hardness-p4}, 
    to prove that even when restricted to networks on $5$ leaves,
    \mcmclong is NP-hard.}
    \label{fig:np-hardness-p4}
\end{figure}
%\end{toappendix}

We still reduce from \setsplit, but proceed a bit differently.
This time, $\N_2$ is a path on $4$ nodes $a,b,c,d$, to which
are respectively connected leaves $\ell_1$, $\ell_2$, $\ell_3$ and $\{\ell_4,\ell_4'\}$ (see Figure~\ref{fig:np-hardness-p4} for an illustration). 
As for $\N_1$, it is composed of 4 nodes $a_1,s,t,b_1$ to which are 
respectively connected $\ell_1,\ell_2,\ell_3$ and $\{\ell_4,\ell_4'\}$.
As previously, two nodes $u_S$ and $v_S$ are introduced
for each $S\in\A$, one node $t_x$ for each $x\in X$. There are edges
from $a_1$ to $s$, from $a_1$ to each $u_S$, from each $u_S$ to the
corresponding $v_S$, from each $v_S$ to $b_1$, from $s$ to $t$,
from $t$ to $b_1$, from $s$ to each $t_x$, and from each $t_x$ to $t$.
Finally, to encode the instance, there are edges from $u_S$ to $t_x$
and $t_x$ to $v_S$ if and only if $x\in S$.
This construction yields the following.

\begin{thm}
    \mcmclong is NP-hard, even on networks with $5$ leaves, 
    one of them being a tree and $k=4$.
    \label{thm:hardness-p4}
\end{thm}
\begin{proof}
    We prove that a \setsplit instance $(X,\A)$ is splittable
    if and only if, with the construction detailed above and 
    illustrated on Figure~\ref{fig:np-hardness-p4}, 
    $\N_2$ is a contraction of $\N_1$. This is equivalent to asking
    whether $(\N_1,\N_2,k)$ is a yes-instance of \mcmc with $k=4$.

    \textbf{$(X,\A)$ splittable $\Rightarrow $ $\N_2$ is a contraction of $\N_1$.}
    Given $X_1,X_2$ split for $(X,\A)$, we describe a sequence of edge contractions
    on $\N_1$ yielding $\N_2$. We first contract $\{t_x\mid x\in X_1\}$
    into $s$ and $\{t_x\mid x\in X_2\}$ into $t$. These
    contractions are admissible, as no other directed path
    from $s$ to $t_x$ other than edge $(s,t_x)$ exist (Proposition~\ref{prop:admissible}). Then, since $\forall S\in A$,
    $S\cap X_1\neq \emptyset$, there is now an edge from $u_S$ to $s$, which
    we contract (one can easily check that it is also admissible). Likewise, we contract each $v_S$ into $t$. The result is $\N_2$.

    \textbf{$\N_2$ is a contraction of $\N_1$ $\Rightarrow $ $(X,\A)$ splittable.}
    By Proposition~\ref{prop:witness_structure}, $\N_2$ is a contraction
    of $\N_1$ if and only if there exists an $\N_2$-witness-structure in $\N_1$,
    i.e.\,a partition $V_a,V_b,V_c,V_d$ of the internal nodes of $\N_1$,
    such that: (1) each of these sets is weakly connected, (2) the parent of $\ell_1$ (resp. $\ell_2,\ell_3,\ell_4,\ell_4'$) in $\N_1$ is in $V_a$
    (resp. $V_b,V_c,V_d$) and (3) there are edges from $V_a$ to $V_b$,
    $V_b$ to $V_c$ and $V_c$ to $V_d$, but all other edges in $\N_1$ are
    internal to each set.
    With such a partition, we must have $a_1\in V_a$, $s\in V_b$,
    $t\in V_c$ and $b_1\in V_d$. If any $u_S$ is in $V_a$, then the
    undirected distance from $V_a$ to $V_d$ is at most $2$ because $u_S$ is
    at undirected distance $2$ from $b_1$ in $\N_1$ and contractions cannot
    increase that distance, whereas that distance must be $3$.  Therefore,
    all the non-leaf neighbors of $a_1$ are in another witness set, and
     $V_a=\{a_1\}$.  By the same argument, no $v_S$ can be in
    $V_d$ and $V_d=\{b_1\}$.  From a similar distance argument, if $u_S\in
    V_c$ then the distance between
    $V_c$ and $V_a$ is incorrect, and we get $\{u_S\}_{S\in\A}\subseteq V_b$ and
    $\{v_S\}_{S\in\A}\subseteq V_c$. 
    Each $t_x$ can only be in $V_b$ or $V_c$. We define $X_1=\{x\in X\mid t_x\in
    V_b\}$ and $X_2 = \{x\in X\mid t_x\in V_c\}$.
    If for some $S\in \A$, $S\cap X_1=\emptyset$, then $u_S$ is not connected
    to the other nodes in $V_b$. Therefore $\forall S\in \A$,
    $S\cap X_1\neq \emptyset$, and likewise for $X_2$. $X_1,X_2$ is indeed
    a split for $(X,\A)$.
\end{proof}

\section{On weakly galled-trees, clades, and contraction sequences}
\label{sec:algo_preliminary_work}

%\ifcomplete
%\begin{toappendix}
\begin{figure}
    \centering
    \includegraphics[width=.9\textwidth]{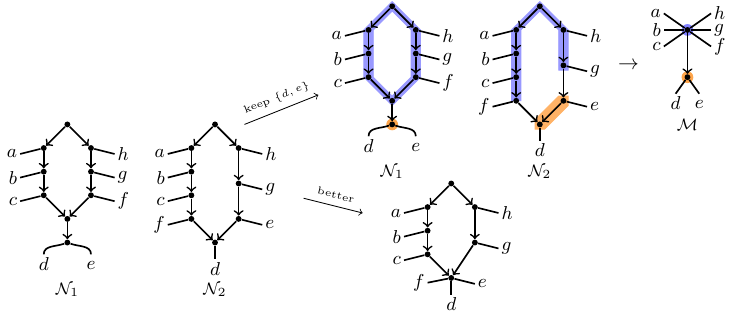}
    \caption{An example showing that not all common $D(u)$ sets should be preserved.  Since $f$ is ``misplaced'', keeping $\{d, e\}$ enforces moving $f$ through the long ``upwards'' path.}
    \label{fig:keep-de}
\end{figure}
%\end{toappendix}

Given the hardness of finding maximum contractions for general networks, we searched for classes of networks that could be solved in polynomial time.  We found that weakly galled trees, defined below, formed such a class.  Our algorithm is slightly involved though, so before diving into it, we explain the main obstacle to overcome when designing algorithm for specific classes.  We then describe a set of reduction rules that can be used to simplify instances.

Recall that for $u \in V(\N)$, we write $D_{\N}(u)$ for the \emph{clade} of $u$, i.e., the set of leaves of $\N$
reached by $u$ (we
recommend~\cite{hellmuth2023clustering} for extensive results on the structural
properties of such clusters).  In the case of two trees $T_1, T_2$, the maximum
contraction subtree is easy to construct: for each $v \in V(T_1)$ such that
$\forall w$ node of $T_2$, $D_{T_2}(w) \neq D_{T_1}(v)$, we contract the edge between $v$ and its parent
in $T_1$.  This algorithm works for three reasons, among others: (1) after
contracting an edge from $T_1$, the result $T_1'$ is still a tree and the algorithm can proceed with the new instance; (2) the set
of clades of $T_1'$ is a subset of the clades of $T_1$; and (3) a clade shared
by both trees is always in the maximum contraction subtree.  Although these
properties may appear trivial, they allow the algorithm to stay applicable
after any number of contractions.

This is not as straightforward in the case of networks.  For (1), if we start
from a galled tree (in which no two cycles contain a common node), a contraction may
result in a network outside of the initial network class (see~\cite{rossello2009all} for precise
definitions).  As for (2), a contraction may create new clades.  Consider
$\N_2$ in Figure~\ref{fig:keep-de}.  There is no $u \in V(\N_2)$ such that
$D_{\N_2}(u) = \{d, e, f\}$.  However, if we contract one of the edges incoming
to the parent of $d$, resulting in $\N_2'$, we create a node $w$ satisfying
$D_{\N_2'}(w) = \{d, e, f\}$.  This new clade is actually part of a maximum
contraction with $\N_1$.  Then for (3), again in Figure~\ref{fig:keep-de},
there are $u \in V(\N_1)$ and $v \in V(\N_2)$ satisfying $D_{\N_1}(u) =
D_{\N_2}(v) = \{d, e\}$.  If we insist on keeping $\{d, e\}$ as a clade, we
must contract large portions of the cycles as shown on the top-right, which is
suboptimal as shown on the bottom-right.  

\paragraph{Weakly galled trees.}

To address point (1) above, 
the so-called \emph{weakly galled trees} appear the most appropriate to study, as they form the simplest class that is closed under contractions, to our knowledge.  We define this class and extend the notion of clades to it.  %A bit more notation is needed.

A \emph{reticulation cycle} of a network $\N$, or \emph{cycle} for short, is a set of nodes $C$ formed by a pair of directed paths $r \rightarrow u_1 \rightarrow \ldots \rightarrow u_p \rightarrow t$ and $r \rightarrow v_1 \rightarrow \ldots \rightarrow v_q \rightarrow t$ that only intersect at nodes $r$ and $t$, and such that $t$ is the only node of $C$ with two in-neighbors from $C$.  We call $r$ the root of $C$, $t$ its reticulation, and the other nodes its internal nodes.
A network is a \emph{weakly galled tree} if 
no pair of reticulation cycles
have an edge in common.  In~\cite{rossello2009all}, the following properties were shown on weakly galled trees: \begin{itemize}
    \item
every reticulation of $\N$ has an in-degree of exactly $2$, and is the reticulation of exactly one cycle;
    \item 
    every cycle $C$ contains exactly one reticulation node of $\N$.  
\end{itemize}
In other words, there is a 1-to-1 correspondence between cycles and reticulations.  Note that the internal nodes of such a cycle $C$ have exactly one child in $C$.
%, with all other children outside of $C$.
We note that a node $v$ could be the root of multiple cycles (which all intersect only at $v$), and that $v$ could be internal to a cycle, and also be the root of another cycle.  The closure property follows from the undirected version of this class, namely cactus graphs, see~\cite{agrawal2021paths}. 
\medskip
\begin{prop}\label{prop:wgt-closure}
    If $\M$ is a contraction of a weakly galled tree $\N$, then $\M$ is also a weakly galled tree.
\end{prop}

\begin{proof}
    It is sufficient to argue that the statement holds if $\M$ is obtained from a single contraction on $\N$, as further contractions will inductively preserve membership to the network class. 
    In~\cite[Observation 5]{agrawal2021paths}, the authors show that if an undirected graph $G$ is a cactus, then contracting any edge also yields a cactus, where a cactus is an undirected graph in which no two cycles share an edge.  In our case, the underlying undirected graph of a weakly galled tree $\N$ is a cactus.  After contracting an edge of $\N$, the underlying undirected graph must thus be a cactus.  Therefore, no two cycles share an edge in this undirected graph, which also holds after orienting the edges, and thus the resulting network is a weakly galled tree. 
\end{proof}

One can verify that a contraction $c(u, v, w)$ cannot increase the number of reticulations.
%(if $z \notin \{u, v\}$ has a single parent, a contraction cannot change this, and $w$ is a reticulation only if one of $u$ or $v$ was a reticulation).  
Therefore, by Proposition~\ref{prop:wgt-closure}, the number of cycles in a
weakly galled tree can only decrease when applying a contraction.  
In fact, the only way to remove a cycle
$C$ is if it has three nodes, that is, a single internal node $x$ that
gets contracted with the root or reticulation. 

%\bm{To give some intuition of weakly-galled trees, we also prove the following useful 
%property.}
%\begin{prop}
%    Let $C_1$,$C_2$ be two cycles of a weakly galled tree $\N$, and $D_1,D_2$ the sets
%    of leaves reachable from an internal vertex of $C_1,C_2$ respectively. If $D_1\cap D_2\neq \emptyset$, then $C_1=C_2$
%\end{prop}
%\begin{proof}
%    
%\end{proof}
%In other words, the sets of leaves reachable from two different cycles can only
%behave in two ways in a weakly galled tree: either one is included in the other, 

%\ifcomplete
%~
%\else 
%Another difficulty of working on networks is that in trees, the sets $D(u)$ of clades in a contraction are contained in the set of clades of the original tree.  In other words, a contraction in a tree $T$ cannot create a clade that did not exist before. 
%This is not true even in weakly galled trees, as new clades can be created by contraction (see Figure~\ref{fig:one_clade_not_kept} in Appendix).  For our algorithms, it is necessary to design an appropriate notion of clade that avoids this problem.
%\fi 

\paragraph{1-clades and 2-clades}

To overcome the difficulties regarding clades mentioned above, 
we next distinguish two types of clades in a weakly galled tree $\N$.  
\begin{itemize}
    \item 
\emph{(1-clades)} A node $u \in V(\N)$ is a \emph{1-clade node} if $u$ is not an internal node of a cycle, in which case $D(u)$ 
is called a \emph{1-clade}. 
\ifcomplete 
Note that a 1-clade node $u$ can be the root of a cycle, but only if $u$ is not internal to another cycle.
\fi
\item 
    \emph{(2-clades)} Let $C$ be a cycle of $\N$ with reticulation $t$.
A pair of distinct nodes $u, v$ of $C$ is called a \emph{2-clade pair} if $u$
and $v$ are on distinct paths that form the cycle, or if one of $u, v$
is the reticulation of $C$ and the other is internal.  The set $D(u)
\cup D(v)$ is called a 2-clade.
\end{itemize}
Note that if $u$ does not belong to a cycle, or is a reticulation, or is a root
of a cycle while not being internal to another cycle (which may happen in weakly galled trees), then $u$ is a 1-clade node.
Also note that multiple nodes may represent the same 1-clade, for instance a
reticulation $u$ with a single child $v$.
Examples of 1-clades and 2-clades may be found on Figure~\ref{fig:mcc_ce_diff}:
$\{1,2,3,4\}$ is a 1-clade of $\N_1$, whose 1-clade node is the root of the cycle
of $\N_1$, and $\{1,2,3\}$ is a 2-clade of $\N_2$, with the parents of $1$ and $3$
as a 2-clade pair. 
%\ifcomplete
Observe that in Figure~\ref{fig:keep-de}, $\{d,e,f\}$ is not a 1-clade of
$\N_1$, but is a 2-clade if we take $u, v$ as the parent of $f$ and the
reticulation, thereby addressing the aforementioned problem with standard
clades. 
%\fi
We let $\Sigma_1(\N)$ denote the set of all 1-clades of $\N$, and
$\Sigma_2(\N)$ denote the set of all 2-clades.  These new types of clades are, in some sense, 
preserved under contractions.

\medskip
\begin{prop}[clade-set conservation]
    If $\M$ is a contraction of a network $\N$, then $\Sigma_1(\M)\subseteq\Sigma_1(\N) \cup \Sigma_2(\N)$ and $\Sigma_2(\M) \subseteq \Sigma_2(\N)$.
    \label{prop:clade_set_conservation}
\end{prop}
\begin{proof}
      It is enough to show that the statement holds if $\M$ is obtained from $\N$ by applying a single contraction $c(u, v, w)$, as further contractions will also  produce networks whose clade-set is a subset of $\M$.  
    So suppose that $\M$ is obtained by contracting $(u, v)$ into the new node $w$.  Note that since $w$ inherits the out-neighbors of $u$ and $v$, we have $D_M(w) = D_\N(u) \cup D_\N(v) = D_\N(u)$ (the latter because $u$ reaches every leaf that $v$ reaches).  
        Let $S$ be a $1$-clade of $\M$. We have $S=D_\M(x)$ for some $x$ $1$-clade
        node of $\M$. 

        If $x\neq w$, then $x$ is also a node of $\N$,
        different from $u$ and $v$. We argue
        it is also a $1$-clade node in $\N$. Indeed, if it is not, then
        it is internal to a cycle in $\N$. Since $x$ is not $u$ nor $v$, no edge incident to $x$ is contracted,
        and $x$ is still part of a cycle in $\M$ (to see this, note that the only way to remove a cycle of $\N$ is if it has a single internal node which is part of the contraction).  Moreover, it is not hard to see that $x$ is still internal to that cycle after contracting, which is a contradiction. Therefore
        $D_\N(x)$ is a $1$-clade of $\N$. 
        This means that in $\N$, $x$ either reaches both $u, v$, or none of them (if not, $x$ would reach $v$ but not $u$, in which case $v$ would be a reticulation and $x$ would be internal to the cycle with that  reticulation).
        This implies that $D_\N(x) = D_\M(x) = S$, as the contraction does not alter the leaves that $x$ reaches.

        If $x=w$, then $S=D_\M(x)=D_\M(w)=D_\N(u)$. Either $u$ was a $1$-clade
        node of $\N$, and $S$ is indeed a $1$-clade of $\N$ and we are done, or $u$ was
        internal to a cycle in $\N$.  Consider the latter case.  
        If $v$ was the reticulation of the cycle, then $u, v$ was a $2$-clade pair of $\N$, with 
        $D_\N(u)\cup D_\N(v)=S$. In that case, $S\in \Sigma_2(\N)$.  If $v$ was not the reticulation, then both $u, v$ were internal to a cycle and one can easily see that $w$ is also internal, and thus not a 1-clade.  It follows that every 1-clade of $M$ is a 1-clade or 2-clade of $\N$.

        Let us now look at the other case, where $S$ is a $2$-clade
        of $\M$. Let $x,y$ be the corresponding $2$-clade node pair,
        and $C$ the cycle they stand on.  
        As cycles may only be deleted,
        and not created, by contractions, $C$ is also a cycle in $\N$,
        possibly with a difference of one edge (if $u,v$ was an edge in $C$).
        Suppose that $w\notin \{x,y\}$, so that $x, y$  are distinct from $u, v$ and thus are also in $\N$.  Then $x, y$ also form a 2-clade pair in $\N$.  If both $x, y$ either reach $u, v$ or none of $u, v$ in $\N$, then the contraction does not change the leaves reached by $x$ and $y$ and our result holds.  
        Suppose that, say, $x$ reaches $v$ but not $u$ in $\N$.  Then $v$ is the reticulation of $C$ and $u$ is its parent on the other side of $x$.  In that case, $y \notin \{u, v\}$ implies that $y$ is an ancestor of $u$ on $C$ and reaches both $u, v$.  We get that $D_\M(x) \cup D_\M(y) = (D_\N(x) \cup D_\N(u)) \cup D_\N(y)$ which is equal to $D_\N(x) \cup D_\N(y)$ since $D_\N(u)$ is included in $D_\N(y)$.  Thus $S \in \Sigma_2(\N)$.

        If $x=w$ or $y=w$, say $x=w$, then we look at a few cases.
        If $w$ is internal to $C$ in $\M$, then through case-checking we see that $u$ was also internal in $C$ in $\N$. 
        Therefore $u,y$ is a
        2-clade node pair, and $S=D_\M(w)\cup D_\M(y)=D_\N(u)\cup D_\N(y)$
        is in $\Sigma_2(\N)$.
        If $w$ is the reticulation of $C$ in $\M$, then either $v$ or $u$
        were the reticulation of $C$ in $\N$ (to see this, expand $w$ and note that either $v$ has $u$ plus another parent, or $u$ receives the parents of $w$). If $v$ was the reticulation,
        then either $u$ was on the same path as $y$ or not. If $u$ was not on the same
        path as $y$, then $u,y$ is a 2-clade node pair in $\N$ and we
        conclude as above. If $u$ was on the same side as $y$, then $D_\N(u)\subseteq D_\N(y)$, and $v,y$ is a 2-clade node pair with clade $S$ in $\N$. 
        Last, if $u$ was the reticulation, then $u,y$ is a valid
        $2$-clade pair node in $\N$, and we have $S=D_\N(u)\cup D_\N(y)\in\Sigma_2(\N).$
\end{proof}

Since new 1-clades or 2-clades cannot be created through contractions, it follows that a 1-clade or a 2-clade that is present in one network $\N_1$, but not in another network $\N_2$, needs to be removed.  In trees, a unique contraction can achieve this, but in networks this is much less obvious.  In the case of 2-clades, we must choose between two possible contractions (one for each side of the cycle) that eliminate the 2-clade.  Even for 1-clade nodes that are the root of a cycle, we may have to choose  between the left or right child to contract.  
Still, we can argue that if we look at clades close to the root, some contractions can be forced unambiguously in a pre-processing step, which is shown very useful in the next section.
Recall that a \emph{reduction rule} alters a given instance if certain conditions are met.  We say that a rule is \emph{safe} if, given networks $\N_1, \N_2$, the rule contracts an edge $(u, v) \in E(\N_1) \cup E(\N_2)$ that is contracted in \emph{any} sequence of contractions leading to a maximum common contraction (note that in the following, the roles of $\N_1$ and $\N_2$ can be swapped to reduce $\N_2$ instead).

\medskip

\noindent 
\textbf{Rule 1.}  If the root $r_1$ of $\N_1$ has a non-reticulation child $u$ such that $D(u)$ is a 1-clade of
$\N_1$, but is not a 1-clade nor a 2-clade of $\N_2$, then contract $(r_1,u)$. 

\medskip

\noindent 
\textbf{Rule 2.}  If the root $r_1$ of $\N_1$ has a child $u$ that is an internal node of a cycle, such that every 2-clade containing $u$ is not a 1-clade
nor a 2-clade of $\N_2$, then contract $(r_1, u)$. 

\medskip

\begin{lemma}\label{lem:rules-are-safe}
    Rule 1 and Rule 2 are safe.
\end{lemma}

\begin{proof}
    Let us argue Rule 1 first.  Let $\M$ be any maximum common contraction of $\N_1$ and $\N_2$.  Note that since $u$ is not a reticulation, there is only one path from $r_1$ to $u$ and contracting $(r_1, u)$ is therefore admissible.  If $(r_1,
    u)$ is never contracted to transform $\N_1$ into $\M$, then $D_{\N_1}(u)$ must be a
    1-clade of $\M$.  Since $\M$ is a contraction of $\N_2$, by Proposition~\ref{prop:clade_set_conservation},
    $D_{\N_1}(u)$ is also a 1-clade or a 2-clade of $\N_2$,
    contradicting the conditions of the rule.

    Let us now show that Rule 2 is safe.  Let $\M$ be a maximum common contraction of $\N_1$ and $\N_2$. 
    We use the witness structure interpretation of $\M$.  That is, let $W_1, W_2, \ldots, W_p$ be the partition of $V(\N_1)$, and $W'_1, \ldots, W'_p$ the partition of $V(\N_2)$ that result in $\M$.  Assume without loss of generality that $r_1 \in W_1$.
    If $u \in W_1$ as well, then $\M$ can be reached after
    contracting $(r_1, u)$ and the rule is safe.  So assume that $r_1 \in W_1$ and $u \in W_2$, again without loss of generality.  

    Let $C$ be the cycle of $\N_1$ in which $u$ is internal.  Let $r_1 = v_1,
    v_2, \ldots, v_k$ be the path of $C$ that does not contain $u$, where $v_k$
    is the reticulation of $C$. We claim that
    $v_k \notin W_1$.  
    Indeed, if $v_k \in W_1$, there are two paths from $r_1$ to $v_k$ that can be used to connect them in $W_1$, and the one that uses $u$ cannot be used.  If we had $r_1, v_k \in W_1$, after contracting all nodes in the same $W_i$'s, 
    there would be an edge from the node representing $W_1$ to that representing $W_2$, because of the edge $(r_1, u)$, and the latter node representing $W_2$ would then have a path to the node representing $W_1$, because $v_k \in W_1$.  This creates a cycle in $\M$ and thus a contradiction.
    Hence, there is some $i \in [k - 1]$ such that
    $v_i \in W_1$ but $v_{i+1} \notin W_1$.  This implies that
    the edges $(r_1, u)$ and $(v_i, v_{i+1})$ of $\N_1$ are not contracted to
    obtain $\M$, implying in turn that $D_{\N_1}(u) \cup D_{\N_1}(v_{i+1})$ is a 1-clade or a
    2-clade of $\M$.  By Proposition~\ref{prop:clade_set_conservation}, this is also
    a 1-clade or 2-clade of $\N_2$, which contradicts the conditions of the
    rule.
\end{proof}

\begin{figure}
    \centering
    \includegraphics[width=.9\textwidth]{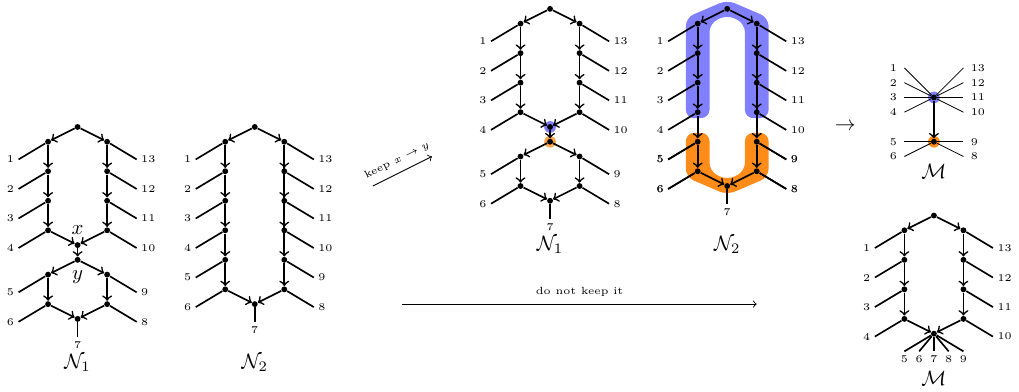}
    \caption{Example of an instance where keeping a common $1$-clade is not optimal. 
    Indeed, if the edge incoming to the root of the second
    cycle in $\N_1$ is kept (corresponding to the 1-clade $\{5,6,7,8,9\}$),
    then its two ends belong to two different sets $W_u$ and $W_v$ of an
    $\M$-witness structure, with $\M$ common contraction of $\N_1$ and
    $\N_2$. One of them, say $u$, must reach exactly $\{5,6,7,8\}$ in $\M$, and $W_u$ must be weakly connected. The only possibility to achieve
    this in $\N_2$ is to contract the cycle into a single edge, as depicted by
    the orange and blue shaded areas on the top path.  Without forcing the
    %keeping 
    preservation of $\{5,6,7,8\}$ (bottom path), we get a larger common
    contraction. 
    This example highlights the difference between computing a
    maximum common contraction of two trees (where one
    can simply keep the clades in common) and two networks.
    }
    \label{fig:one_clade_not_kept}
\end{figure}

Note that if, for a 1-clade node $u$ of $\N_1$, the set
$D(u)$ is in $\N_2$ 
%\st{as either a 1-clade or} 
as a 2-clade,
%\bm{if it is two 1-clades they are kept}
we may still have to
contract the edge between $u$ and its (unique) parent.  Hence, an algorithm to find common contractions cannot simply keep the common 1-clades and 2-clades, as is done on trees.  An example
is given in Figure~\ref{fig:one_clade_not_kept}.
A common 2-clade may also need to be removed.

%\begin{toappendix}
%\end{toappendix}

\paragraph{Unicity of clades} To finish this section, we prove a useful Lemma
about the limited number of ways a set of leaves may be represented as a 1-clade or 2-clade
in a weakly galled trees. We show that this number is constant, which will be 
used in the analysis of our algorithm, in the next section. In its proof, we use
the following Claim:

\medskip
\begin{lemma}
    Given a weakly galled tree $\N$ without degree-$2$ nodes and a set of leaves $S$, there is:
    \begin{itemize}
        \item[1.] At most two 1-clade nodes $u$ such that $D(u)=S$
        \item[2.] At most one 2-clade pairs $u,v$ such that $D(u)\cup D(v)=S$ 
    \end{itemize}
    \label{lemma:clade_unicity}
\end{lemma}
\begin{proof}
    Assume that there are distinct nodes $u, v$ that are 1-clade nodes such that $D(u) = D(v) = S$.  This is possible if $v$ is the unique child of $u$, so assume this is not the case. 
    Suppose that $u, v$ are incomparable, that is, none is a descendant of the
    other.  Let $P$ be any path from $u$ to a leaf $\ell$.  Because $D(v) =
    D(u)$, $v$ also reaches $\ell$, and any path $P'$ from $v$ to $\ell$ must
    intersect $P$ because, in particular, both $P$ and $P'$ must contain the
    parent $p(\ell)$ of $\ell$.  The first node $t$ at which $P$ and $P'$
    intersect is a reticulation since it receives a distinct in-neighbor from
    both paths.  Note that $t \notin \{u, v\}$ since $u, v$ are incomparable and
    they both reach $t$.  
    Moreover, $u$ and $v$ must have a common ancestor $w$ (in particular, the
    root is such an ancestor).  This means that $u$ and $v$ are part of some
    cycle whose reticulation is $t$.  By incomparability, $u$ and $v$ are
    internal to that cycle, and thus cannot be 1-clade nodes, a contradiction.  

    Next suppose that $u$ and $v$ are comparable, and without loss of generality
    that $u$ reaches $v$.  If $u$ has only one child $w$, then $u$ is a
    reticulation or the root and under our current assumption (that $v$ is not the only child of $u$) 
    we have that $w
    \neq v$.  Also, $D(u) = D(w)$ and $w$ cannot be a reticulation (if it was,
    its in-neighbors would be on the cycle of $w$, including $u$ --- as the
    latter is a 1-clade, $u$ would root  that cycle and have two children). 
    In that case, redefine $u$ to be $w$ ($v$ still descends from $u$).  In that manner, we may assume that $u$ has at least two children $p$ and $q$.  If $u$ is not the root of any cycle, then $D(p)$ and $D(q)$ are disjoint, and, as $v$ descends from $u$, $D(v)$ cannot contain both $D(p)$ and $D(q)$.  So assume that $u$ is the root of some cycle $C$ that contains $p$ and $q$ and a reticulation $t$.  Note that $v$ is not internal on $C$ as it is 1-clade.  If $v \neq t$ or if $v$ does not descend from $t$, then $t$ must reach leaves that $v$ does not.  So we may assume that $v$ is $t$ or descends from it.  In that case, assume without loss of generality that $q$ is internal to $C$.  Then $q$ has some child $q'$ not on $C$.  Moreover, $q'$ reaches some leaf that $t$, and thus $v$, does not reach.  Thus $D(u) \neq D(v)$, and therefore only two nodes can have the same 1-clade. 

    As for Point 2, let $u,v$ be a 2-clade pair such that $D(u)\cup D(v)=S$.
    Let us call $C$ the cycle they belong to, $r$ the root of $C$ and $t$
    its reticulation.
    Consider $w,x$ another 2-clade pair such that $D(w)\cup D(x)=S$, belonging to cycle $C'$ with root $r'$ and reticulation $t'$. We will show that $w,x$
    can only be equal to $u,v$. 

    Recall that in a 2-clade pair, at least one vertex must be internal to its
    cycle.  Say that $u$ and $w$ are internal, and thus have a child outside of
    their respective cycles.  Suppose that $C \neq C'$, and thus $t \neq t'$.
    If $t$ is an ancestor of $t'$, the child of $u$ outside of $C$ must reach a
    leaf that $w$ and $x$ cannot reach (otherwise, there would be a cycle with
    two reticulations from $u$ going through $t$ then $w$ or $x$, then to the
    intersection on the paths leading to that leaf). A similar argument applies
    if $t'$ is an ancestor of $t$.  
    
    If $t, t'$ are incomparable, we first show that $D(t)\cap D(t')=\emptyset$.
    Indeed, suppose $D(t)\cap D(t')\neq \emptyset$ and let $\ell$ be a leaf
    in that intersection.   
    As in the proof of Point 1 above, let $P$ be any path from $t$ to $\ell$, and
    $P'$ any path from $t'$ to $\ell$. These two paths intersect at least at the parent of $\ell$.
    The first node $s$ at which they intersect must be a reticulation (as if it had only one
    in-neighbor, this node would be in both $P$ and $P'$, and $s$ would not be the first intersection).
    In addition, $t$ and $t'$ also have a common ancestor (the root of the network for instance).
    The lowest common ancestor of $t$ and $t'$ is therefore the root of a cycle whose reticulation
    is $S$. This is not possible in a weakly galled network, in which reticulations (such as $t$ and $t'$)
    belong to exactly one cycle. Therefore if $t$ and $t'$ are incomparable, $D(t)\cap D(t')=\emptyset$.
    We show that this implies that the roots $r$ and $r'$ of $C$ and $D'$ must also be incomparable.
    Indeed, if $r$ can reach $r'$ while $t$ and $t'$ are incomparable, then 
    %it is either because $r'$
    %is internal to $C$, or $r'$ is reachable from an internal node of $C$ using one o its edges outside
    %of the cycle (which must exist if there are no degree-2 nodes).
    either $r'$ is on the cycle $C$, in which case $C, C'$ intersect only at $r'$ (here $r'$ could be equal to $r$ or internal to $C$), or $r'$ is reachable from $r$ by going to a node of $C$ (possibly $r$) then following a path of edges not on $C$.
    In any case, there is a leaf
    in $D(t)$ (and therefore in $D(u)\cup D(v)$) unreachable by both the
    internal vertices or the reticulation of $C'$, and in particular $w$ and
    $x$. This is a contradiction. A similar argument applies if $r$ is reachable from $r'$.
    However, if $r$ and $r'$ are incomparable, then by the same argument applied to $t$ and $t'$
    (taking a leaf reachable by both, and a common ancestor), we get that $r$ and $r'$ are part of the same cycle,
    which is a contradiction.

    %Moreover, $C$ and $C'$ either do
    %not intersect, or intersect at their roots \bm{(i.e.\,either they share the same root, 
    %or the root of one of the cycles belongs to the other)}. 
    %[TODO FINISH THIS]. 
    Therefore, if $C\neq C'$, in all cases, $D(w) \cup D(x)$ cannot be equal to $D(u) \cup D(v)$.  
    So we may assume that $t = t'$ and $C = C'$.  
    Suppose, without loss of generality, that among $u, v, w, x$, the node $u$ is at minimum distance from $r$ on $C$.  
    If $u \notin \{w, x\}$, then the child of $u$ outside of $C$ reaches a leaf
    that neither $w$ nor $x$ can reach, because $w, x$ are either below $u$ or
    on the other side of $C$. So assume $u = w$.  Then consider the node among
    $v, x$ that is the closest to $r$, say $v$.  If we assume $v \neq x$, then
    $v$ must be internal in $C$, and the child of $v$ outside of $C$ must reach
    a leaf that $x$ cannot reach (because $x$ descends from $v$ on $C$) and that
    $w$ cannot reach (because it is on the other side of the cycle).  Therefore,
    $\{u, v\} = \{w, x\}$.

\end{proof}

%\begin{lemma}
%Let $\N$ be a weakly galled tree  and $C$ a cycle
%of $\N$, whose root is $r$. If $v$ is any node of $\N$ not in $C$, then 
%either $D(r)\subseteq D(v)$, $D(v)\subseteq D(r)$ or $D(r)\cap D(v)=\emptyset$.
%\label{lemma:disjointness}
%\end{lemma}
%\begin{proof}
%   If $r$ is reachable from $v$, then $D(r)\subseteq D(v)$. Suppose therefore
%   $r$ is not reachable from $v$. We will show that in this case, $D(r)\cap
%   D(v)=\emptyset$.
%
%   Indeed, suppose a leaf $\ell$ is reachable from both $v$ and $r$.
%\end{proof}
%
%A consequence of Lemma~\ref{lemma:disjointness}, which will be useful in the next section,
%is that if a node $u$ contains \emph{some} of the leaves descending from a cycle,
%but not all, then it must be internal to the cycle.
%}

\section{A Dynamic Programming Algorithm for Weakly Galled Trees}
\label{sec:algo}

We now show that computing the minimum number of contractions needed to achieve a common contraction for a pair of weakly galled trees $\N_1, \N_2$ is feasible in polynomial time.  
Let $C$ be a cycle of $\N \in \{\N_1, \N_2\}$ with root $r$.  It will be convenient to use a cyclic ordering of its nodes.  
Let $r, v_1, v_2, \ldots, v_{l}, r$ be the sequence of nodes obtained by traversing the cycle $C$ when seen as an undirected graph, starting at $r$ and choosing one of its neighbors $v_1$ arbitrarily.  We view this as traversing the cycle counter-clockwise. 
For a node $u$ of $C$, we write $u + 1$ for the node that succeeds $u$ in this sequence, and $u - 1$ for the node that precedes $u$ (note that this is also well-defined for $r$, as it has only one predecessor, which is $v_\ell$, and one successor, which is $v_1$).  For $u , w$ in $C$, we write $u \leq_C w$ if $u = w$ or $u$ precedes $w$ in the sequence, and we drop the $C$ subscript if clear.
As a special case, for every $u$ of $C$ both $u\leq_C r$ and $r\leq_C u$. %$r+1=v_1$ and $r-1=v_\ell$. %$u \leq r$ for every $u$ of $C$.
For $u \leq v$, we denote by $|v - u|$ the number of edges on the path from $u$ to $v$, following the cyclic ordering.  If $v < u$, then $|v - u| = 0$.

%\vspace{-3mm}

\paragraph*{Some useful subnetworks}

\begin{figure}
    \centering
    \includegraphics[width=.99\textwidth]{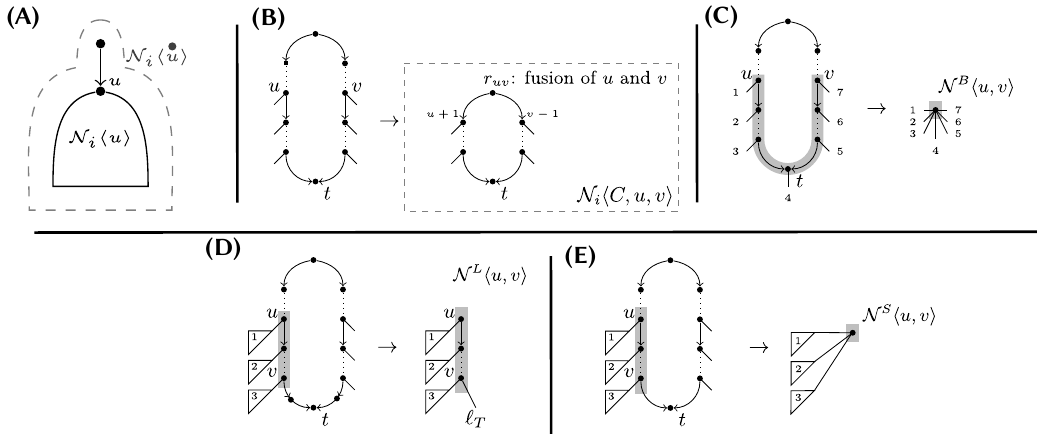}
    \caption{Sub-networks used in our polynomial algorithm for weakly galled
    trees. The dynamic programming is primarily based on
    $\N_i\langle\dang{u}\rangle$ and $\N_i\langle C,u,v\rangle$ ((A) and (B))
    on the figure. The others are intermediate networks on which
    reduction rules are applied to fall back to 
    $\N_i\langle\dang{u}\rangle$ and $\N_i\langle C,u,v\rangle$.
%    \ml{[todo: swap the position of $\N^S$ and $\N^L$]}
}
    \label{fig:subnetworks}
\end{figure}

Let $\N$ be a weakly galled tree.  In our recurrences, we will make use of the following networks that can be obtained from $\N$. They are all displayed
in Figure~\ref{fig:subnetworks}.
\begin{itemize}
\item 
    $\N\langle u\rangle$ and $\N \langle \dang{u}\rangle$
    (Figure~\ref{fig:subnetworks}~(A)): for $u$ a 1-clade
    node, $\N\langle u\rangle$ is the network induced by $u$ and all the
    nodes that $u$ reaches. $\N \langle \overset{\bullet}{u}\rangle$ is obtained from $\N \ang{u}$ by adding a new node $r$ and the
    edge $r \rightarrow u$.  The circle above $u$ represents a
    ``dangling'' root with a single child $u$.  
%        
%
%        \begin{center}
%            \includegraphics[width=.2\textwidth]{tikz_figures/one_clade_subnetwork.pdf}
%        \end{center}
%    
%
%
\item 
    $\N \ang{C, u, v}$ (Figure~\ref{fig:subnetworks}~(B)): for $C$ a cycle of $\N$ with reticulation $t$, and $u < t < v$, $\N \ang{C, u, v}$ is 
    obtained from the network induced by any node reached by $u + 1, v - 1$ (inclusively), then adding a new node $r_{uv}$ with children $u + 1, v - 1$. 
    %\ots{[Suggestion: add $r_{uv}$ and maybe $u + 1, v - 1$ in Fig 4 B]}.  
    Such a sub-network can be seen as the result of contracting the upper part of the cycle between $u$ and $v$ into its root.  
    
    It is possible that $u + 1 = v - 1$, in which case they are the reticulation $t$ of the cycle.  In this case, notice that $\N \ang{C, u, v}$ is equal to $\N \ang{\overset{\bullet}{t}}$.
    It is also possible that $u = v$ is the root of $C$ (which is why we need to specify $C$ in the notation, since $u$ can root multiple cycles).  In this case, the cycle $C$ is kept as is, but if $u$ has children outside of $C$, they are removed.
%    
%
%\begin{center}
%    
%    \includegraphics[width=.2\textwidth]{tikz_figures/two_clade_subnetwork.pdf}
%
%        \end{center}
%    
%
    \item 
    $\N^B \ang{u, v}$ (Figure~\ref{fig:subnetworks} (C)): for $u,v$ a 2-clade node pair, or $u = v$ a reticulation, $\N^B \ang{u, v}$
        is the subnetwork obtained by (1) removing any node not descending
        from $u$ or $v$ and (2) contracting the path from $u$ to $v$ going
        through the reticulation of the cycle.  Note that $u$ or $v$ could be the reticulation.  The $B$ stands for \emph{bottom}.
%
%    \begin{center}
%               \includegraphics[width=.4\textwidth]{tikz_figures/bottom_subnetwork.pdf} 
%        \end{center}
%
%
%
    
%
%    \begin{center}
%        \includegraphics[width=.4\textwidth]{tikz_figures/contracted_side_subnetwork.pdf}
%        \end{center}
%
%    
    \item 
    $\N^L \ang{u, v}$ (Figure~\ref{fig:subnetworks}~(D)): for $u,v$ a pair of internal nodes belonging
    to the same side of a cycle, $\N^L \ang{u,v}$ is the network induced by all
    the nodes reached by $u$, but that are not reached by $v + 1$.  
    For technical reasons, an extra leaf $\ell_T$ is added, with $v$ as its parent (which simulates that there existed a lower portion of the cycle). 
    Note
    that this network includes $u$ and excludes $v + 1$.
    Moreover, if $v$ is not $u$ or one if its
    descendants, we assume that $\N^L \ang{u, v}$ is a network with no node.
    The $L$ stands for \emph{lateral}.

    \item $\N^S \ang{u, v}$ (Figure~\ref{fig:subnetworks}~(E)): for $u,v$ pair of nodes belonging
    to the same side of a cycle, $\N^S \ang{u,v}$ is obtained from $\N^L \ang{u, v}$ 
    by removing $\ell_T$ and contracting the path from $u$ to $v$.  If $v$ does not descend from $u$, this is the empty network.  The $S$ stands for \emph{side}.

\end{itemize}

\medskip

\noindent
\textbf{Join subnetworks.}
Let $\N_1, \N_2$ be two networks with distinct leaf sets.  
We write $\N_1 \ast \N_2$ for the network obtained by identifying the roots of $\N_1$ and $\N_2$ (or, to explain this differently, we create a new root that inherits all the children of the roots of $\N_1$ and $\N_2$, and delete the two previous roots).  
If $\N$ is a weakly galled tree, we say that a network $\N'$ is a \emph{join subnetwork of $\N$} if $\N' = \N^1 \ast \ldots \ast \N^p$, 
with each $\N^i$ either of the form $\N \ang{\dang{u}}$ for some $u \in V(\N)$, or $\N \ang{C, u, v}$ for some cycle $C$ of $\N$ and $u,v$ on $C$.  
If $p = 1$, then $\N'$ is called a \emph{prime join subnetwork} of $\N$, and otherwise it is a \emph{composite join subnetwork} of $\N$.  Note that by the definition of $\ast$, no two $\N^i$ subnetworks may have a leaf in common.
Two join subnetworks $\N_1=\N_1^1\ast\dots\ast\N_1^p$ and $\N_2=\N_2^1\ast\dots\ast\N_2^q$ are said to be \emph{matching}
if $p=q$ and $\forall i\in[p]$, $L(\N_1^i)=L(\N_2^i)$. 

Our dynamic programming tables will contain join subnetworks as entries.    The following will be useful to argue that applying contractions keeps us in the realm of join subnetworks.

\medskip
\begin{lemma}\label{lem:contr-maintains-join}
    Let $\N$ be a weakly galled tree and let $\N'$ be a join subnetwork of $\N$.  Then applying to $\N'$ any sequence
    of contractions of edges outgoing from its root yields a join subnetwork of $\N$.
    \label{root_contractions}
\end{lemma}
\begin{proof}
   It suffices to argue that a single contraction preserves the desired form, as in turn, applying any number of them will do so.
    Let $r$ be the root of $\N'$ and suppose that edge $r\rightarrow w$ is contracted.  Denote $\N' = \N^1 \ast \ldots \ast \N^p$.
    Assume without loss of generality that $w$ belongs to $\N^1$.  
    Suppose first that $\N^1 = \N \ang{\dang{w}}$ for some $w \in V(\N)$.  
    Notice that because $\N$ is a weakly galled tree, $\N \ang{w}$ is itself a composite join subnetwork of $\N$, as it can be obtained by joining its 1-clade children under a common root, and joining the cycles that $w$ is a root of.  In other words, we may write $\N \ang{w} = \N_w^1 \ast \ldots \ast \N_w^q$.  Then, after contracting $r \rightarrow w$, the network $\N'$ becomes $\N_w^1 \ast \ldots \ast \N_w^q \ast \N^2 \ast \ldots \ast \N^p$, which is a join subnetwork of $\N$, as desired.
    So suppose that $\N^1 = \N \ang{C, u, v}$ for some cycle $C$ and nodes $u, v$.  Then $w = u + 1$ or $w = v - 1$.  Either way, if the cycle is still present after contracting $r \rightarrow u$, $\N^1$ can be replaced with $\N \ang{u+1, v}$ or $\N \ang{u, v - 1}$, in which case the contraction yields another join subnetwork (as the other subnetworks $\N^2, \ldots, \N^p$ are unaltered).  If the cycle is contracted to an edge, then $\N^1$ becomes $\N \ang{\dang{t}}$, with $t$ the reticulation of the cycle.  Again, this preserves the join subnetwork form.
    \end{proof}

The last but not least ingredient before we can describe our procedure is to use Rules 1-2 from the previous section in order to reach join subnetworks with corresponding leaf sets.

\begin{lemma}\label{lem:reduced-form}
    Suppose that Rules 1-2 are not applicable to $\N_1$ or $\N_2$. 
    Then the networks can be written as matching join subnetworks $\N_1 = \N_1^1 \ast \ldots \ast \N_1^p$ and $\N_2 = \N_2^1 \ast \ldots \ast \N_2^p$ such that
    $L(\N_1^i) = L(\N_2^i)$ for every $i \in [p]$.
\end{lemma}

\begin{proof}
    Let $r_1$ and $r_2$ be the roots of $\N_1$ and $\N_2$, respectively.  
    Notice that a network is a join subnetwork of itself, so we may write 
    $\N_1 = \N_1^1 \ast \ldots \ast \N_1^p$ and $\N_2 = \N_2^1 \ast \ldots \ast \N_2^q$.  If the superscripts can be arranged so that leaf sets are equal as desired in the statement, we are done, so assume this is not the case.
    Let us start with an observation:
\\
    \textbf{Claim.} If two leaves $x$ and $y$ from a network $\N$ belong to a $1$-clade or $2$-clade other than $L(\N)$, then there is an (undirected) path between them that does not
        use the root.\\
        \textbf{Proof of Claim:}
        In case of a $1$-clade $S=D(u) \neq L(\N)$, simply concatenate a path from $x$ to
        $u$ (which must exist as $u$ reaches $x$) and a path from $u$ to $y$.
        Likewise, in the case of a $2$-clade $S=D(u)\cup D(v)$, use $u$ and $v$, then the reticulation of the cycle that $u$ and $v$ belong to,
        as intermediate points to get a path from $x$ to $y$ not using the root.\qed
          
    The matching condition holds if  $\forall i\in[p]$, $\exists j $
    such that $L(\N_1^i)=L(\N_2^j)$. Therefore, if it is not verified,
    there exists $i\in [p]$ such that $\forall j\in[q]$, $L(\N_1^i)\neq L(\N_2^j)$.
    Since $\{L(\N_2^j)\}_{j\in[p]}$ forms a partition of the common leafset
    of $\N_1$ and $\N_2$, there must exist $j$ such that, in addition 
    $L(N_1^i)\cap L(\N_2^j)\neq \emptyset$. 
    Let us pick such a pair $i,j$ and denote $A=L(\N_1^i)$ and $B=L(\N_2^j)$.
    We have $A\neq B$ and $A\cap B\neq \emptyset$, and therefore $B\setminus A\neq
    \emptyset$ or $A\setminus B\neq \emptyset$. We analyze the case $B\setminus A\neq \emptyset$, but the arguments apply symmetrically to the other.
%    \st{Let $i \in [p]$ be the smallest value such that $L(\N_1^i) \neq L(\N_2^j)$, for every $j \in [q]$.  
%    To ease notation, let us denote $A = L(\N_1^i)$.  Let $\N_2^j$ be any
%    element composing $\N_2$ such that $A \cap L(\N_2^j) \neq \emptyset$, noting
%    that this must exist.  Let us denote $B = L(\N_2^j)$.  Note that $A \neq B$,
%    and that} 
    Note that $\N_2^j$ is either rooted at a node with a single child, or rooted
    at a cycle.

    Consider the case where the root $r_2$ of $\N_2^j$ has a single child $v$.
    Let $x\in A\cap B$ and $y\in B\setminus A$.
    We have that $v$ is a child of $r_2$ in $\N_2$ and, moreover,
    $D_{\N_2}(v) = B$ is a 1-clade of $\N_2$ containing both $x$ an $y$.
    However, in $\N_1$, all
    paths between $x$ and $y$ go through $r_1$, and by contraposing
    the Claim above,
    $x$ and $y$ cannot share a clade, implying that $B$ cannot be a 1-clade
    nor a 2-clade of $\N_1$. Therefore, Rule 1 is
    applicable to $r_2 \rightarrow v$, a contradiction.

    So consider the case where $\N_2^j$ is rooted at a cycle $C$.  Let $t$ be the reticulation of $C$ and let $x$ 
    be a leaf that $t$ reaches (in $\N_2$ and $\N_2^j$).  Suppose for now that $x \in A$.  Then let $y \in B \setminus A$, which again exists.  Let $v$ be a non-reticulation child of $r_2$ in $\N_2$ that reaches $y$ (which exists), and note that $v$ also reaches $x$.  Therefore, any 2-clade involving $v$ contains $x$ and $y$.  Again because all paths between $x$ and $y$ in $\N_1$ go through $r_1$, such a 2-clade cannot be in $\N_1$, in which case Rule 2 should be applied to $r_2 \rightarrow v$, a contradiction.  
    So suppose that $x \notin A$.  Then let $y \in A \cap B$.  As before, we let $v$ be a non-reticulation child of $r_2$ that reaches $y$.  This $v$ also reaches $x$, and we get the same contradiction.
\end{proof}

%\vspace{-3mm}

\paragraph*{A dynamic programming algorithm}

\begin{figure}
\centering
\includegraphics[width=\textwidth]{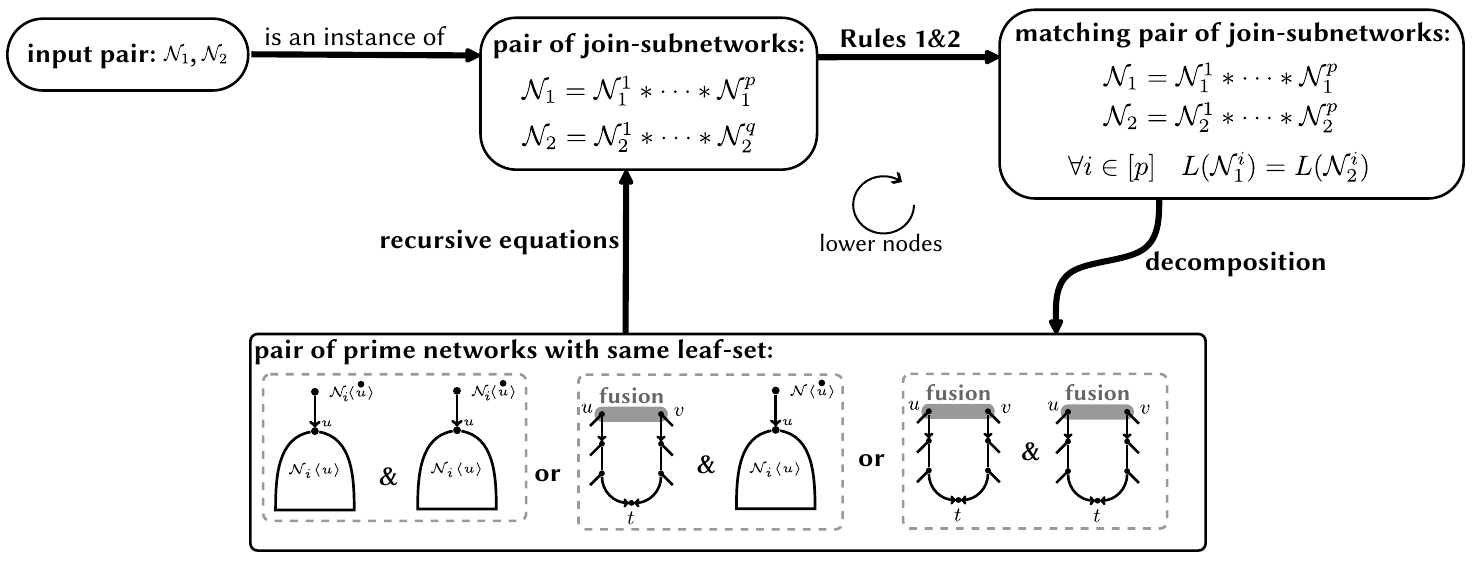}
    \caption{Overview of our algorithm for weakly galled trees. }
    \label{fig:workflow}
\end{figure}

We next describe our dynamic programming algorithm.
Its overall workflow is sketched in Figure~\ref{fig:workflow}, relying on the
concept of join subnetwork. The present section
details the ``decomposition'' and ``recursive equations'' displayed in Figure~\ref{fig:workflow}.

We define two functions $\fp$ and $\fc$, linked together by recursive
equations. They both compute, given two join subnetworks $\N'_1,\N'_2$ of $\N_1$
and $\N_2$, the minimum number of contractions
needed on both networks to achieve a common contraction. The difference
is that $\fp(\N'_1, \N'_2)$ assumes that the given pair are \emph{prime} join subnetworks, whereas
$\fc(\N'_1, \N'_2)$ is given an arbitrary pair of join subnetworks (which may be composite or prime).
The role of $\fc$ is to apply rules 1 and 2, in order to get a matching
pair of join-subnetworks, and then call $\fp$ on the resulting pairs of prime
subnetworks.
Our value of interest is $\fc(\N_1, \N_2)$.

For join subnetworks $\N'_1, \N'_2$, we start with the simple base cases.

\medskip

\noindent 
\textbf{Base cases.}  
If $L(\N_1') \neq L(\N_2')$ or if one of $\N_1', \N_2'$ is not acyclic, then $\fc(\N_1', \N_2') = \fp(\N_1', \N_2') = \infty$ as no common contraction is possible.  
If none of the above holds and if $\N_1'$ and $\N_2'$ each have no node, or each has a single node, put $\fc(\N_1', \N_2') = \fp(\N_1', \N_2') = 0$.

\medskip

\noindent 
\textbf{Main entries $\fc$.}
If $\N'_1$ and $\N'_2$ are both prime join subnetworks, then 
$\fc(\N'_1, \N'_2) = \fp(\N'_1, \N'_2)$ (defined later below).  Otherwise, suppose that at least one of $\N_1'$ or $\N_2'$ is a composite join subnetwork of $\N_1$ or $\N_2$.  
Then proceed as follows:
\begin{itemize}
    \item 
    As long as one of Rule 1 or Rule 2 is applicable to $\N_1'$ or $\N_2'$, we apply it.  Let $\N''_1, \N''_2$ be the resulting pair of networks, which are join subnetworks by Lemma~\ref{lem:contr-maintains-join}, and suppose that Rule 1-2 enforced $k$ contractions.   Then we have $ \fc(\N_1', \N_2') = k + \fc(\N_1'', \N_2'')$

    \item 
    If no rule is applicable to either network, by Lemma~\ref{lem:reduced-form}, we have matching join subnetworks $\N_1' = \N_1^1 \ast \ldots \ast \N_1^p$ and $\N_2' = \N_2^1 \ast \ldots \ast \N_2^p$, with $L(\N_1^i) = L(\N_2^i)$ for each $i \in [p]$.  We can solve each pair of prime subnetworks independently with
        $\fc(\N_1', \N_2') = \sum_{i=1}^p \fp(\N_1^i, \N_2^i)$.
\end{itemize}

\medskip

\noindent 
\textbf{Prime entries $\fp$.}
We now assume that both $\N_1'$ and $\N_2'$ are prime join subnetworks of $\N_1$ and $\N_2$, respectively.  
By definition, $\N'_1$ must have the form $\N'_1 = \N \ang{\dang{u}}$ or $\N'_1 = \N \ang{C, u, v}$.
The same holds for $\N'_2$.  There are thus four combinations to check. 

\medskip

\noindent 
\textbf{Case 1:} For $u \in V(\N_1)$, $v \in V(\N_2)$:
$\fp(\N_1 \ang{\dang{u}}, \N_2 \ang{\dang{v}}) = \fc(\N_1 \ang{u}, \N_2 \ang{v})$.
This is because both networks start with a single edge $r_u \rightarrow u$
and $r_v \rightarrow v$ (respectively). An optimal common contraction
simply keeps both of these edges and defers the computation to
 $\N_1 \ang{u}$ versus $\N_2 \ang{v}$, which are join subnetworks (possibly prime) and can use other $\fc$ entries.

\medskip

\noindent 
\textbf{Case 2:} For $u \in V(\N_1)$, and cycle $C$ of $\N_2$ and $v, w \in V(\N_2)$ belonging to $C$:
\begin{align*}
\fp(\N_1 \ang{\dang{u}}, \N_2 \ang{C, v, w}) &= \min 
    \begin{cases}
        \fc(\N_1\langle u\rangle,\N_2^B\langle
v+1,w-1\rangle)+|w-v|-2,\\ 
        \fc(\N_1 \langle u\rangle,\N_2\langle C, v,w\rangle)+1
    \end{cases}
\end{align*}

Here, $\N_1 \ang{\dang{u}}$ starts with a single edge $r_u\rightarrow u$,
while $\N_2 \ang{C, u, v}$ starts with a root of a cycle. The minimization
distinguishes two cases: either (a) edge $r_u\rightarrow u$ is conserved,
in which case the cycle must be contracted to a single edge, with the root having one child, or (b) the edge is
contracted.  
%\ml{[I have doubts on the first case: $\N_2^B$ contracts to a
%single point, not a single edge.]} 
%\bm{[I think you are right, there should not
%be a point on the first $\N_1$]}

\medskip

\noindent 
\textbf{Case 3:} For a cycle $C$ of $\N_1$, $u, v \in V(\N_1)$ and $w \in V(\N_2)$, the computation of $\fp(\N_1 \ang{C, u, v}, \N_2 \ang{ \dang{w} })$ is symmetric to the previous case.

\medskip

\begin{figure}
    \centering
    \includegraphics[width=0.95\textwidth]{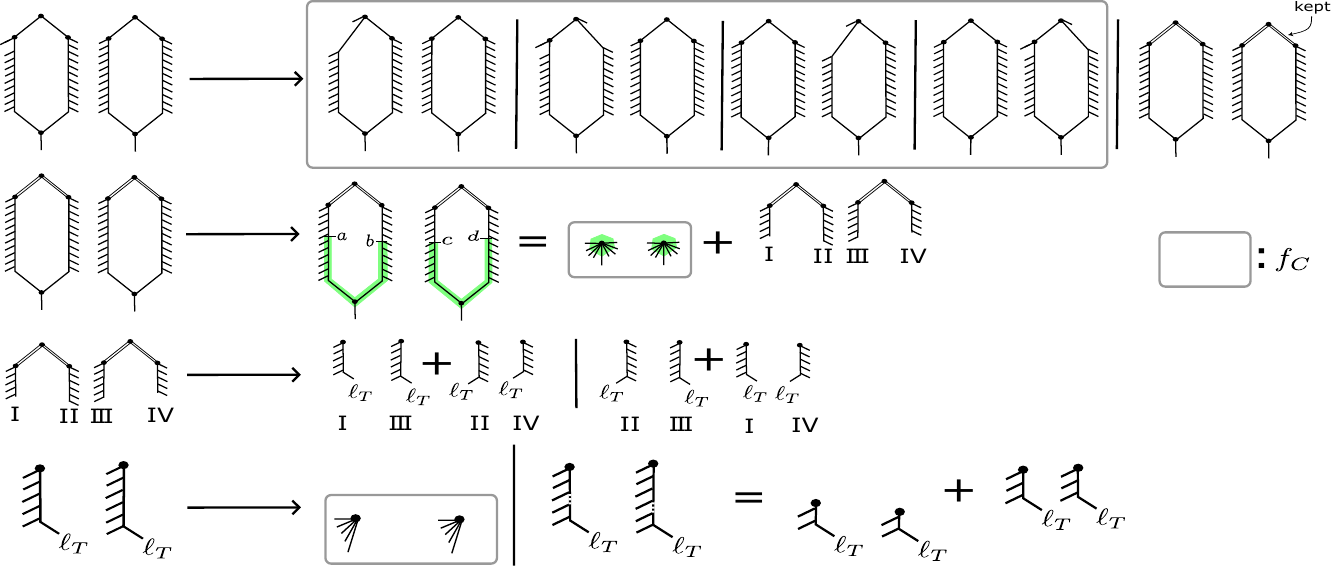}
    \caption{The handling of Case 4.  The gray boxes represent calls to $\fc$ entries, and vertical bars separate possible cases.  
    First row: The left and right children of the first network are $u+1, v - 1$, and the left and right children of the second are $w + 1, x - 1$.  In the gray box, the four ways of contracting an edge incident to a root (in bold edges is the untouched network).  If we do not apply those, we keep these edges.
    Second row: we choose $a,b,c,d$ and contract the corresponding paths and defer the computation to $\fb(a,b,c,d)$, which requires optimizing over the $\N_1^B, \N_2^B$ bottom networks, plus what is above $a,b,c,d$.
    Third row: the two ways of matching the lateral networks, considered by $\fb$.
    Fourth row: given two lateral networks for an $\fl$ entry, we either contract the whole lateral paths to get $\N^S$ networks, or split the paths and solve two smaller lateral networks.}
    \label{fig:case4}
\end{figure}

\noindent 
\textbf{Case 4:} 
This is the most complicated case, which is illustrated in Figure~\ref{fig:case4}.
For a cycle $C_1$ of $\N_1$ and nodes $u, v \in C_1$, and cycle $C_2$ of $\N_2$ and $w, x \in C_2$, with respective cycle roots $r_{uv}, r_{wx}$ and reticulations $t_1, t_2$, let us consider the following possibilities for $\fp(\N_1 \ang{C_1, u,v}, \N_2 \ang{C_2, w, x})$: either we contract an edge incident to $u, v, w$, or $x$ on the cycles, or not.  In the latter case, we treat these incident edges as ``kept'', and we must handle the bottom of the cycles.    For clarity, this value is described in a separate and temporary entry $\fb(a, b, c, d)$, which represents the best scenario if we contract the subpath of $\N_1 \ang{C_1, u, v}$ from $a$ to $b$ into $t_1$, and the subpath of $\N_2 \ang{C_2, w, x}$ from $c$ to $d$ into $t_2$.  We get:
\begin{align*}
    \fp(\N_1 \ang{C_1, u,v}, \N_2 \ang{C_2, w, x}) = \min 
    \begin{cases}
        \fc(\N_1 \ang{C_1, u, v} / (r_{uv}, u+1), \N_2 \ang{C_2, w, x}) + 1 \\
        \fc(\N_1 \ang{C_1, u, v} / (r_{uv}, v-1), \N_2 \ang{C_2, w, x}) + 1 \\
        \fc(\N_1 \ang{C_1, u, v}, \N_2 \ang{C_2, w, x} / (r_{wx}, w + 1)) + 1 \\
        \fc(\N_1 \ang{C_1, u, v}, \N_2 \ang{C_2, w, x} / (r_{wx}, x-1)) + 1 \\
        \min_{\substack{u < a \leq t_1 \leq b < v \\ w \leq c \leq t_2 \leq d \leq x \\
        D(a)\cup D(b)=D(c)\cup D(d)}} \fb(a, b, c, d) \\
    \end{cases}
\end{align*}

Note that the first four entries create a composite join subnetwork, as one root inherits the prime subnetworks going out of its contracted child.  In the event that one of these contractions is inadmissible, a cycle is created and the base case returns $\infty$.  

 An $\fb$ entry needs to consider the contractions at the bottom of the cycle, plus the join subnetworks created at the bottom, plus the contractions needed on the two lateral parts of the cycles.  The latter requires comparing subnetworks of the form $\N_1^L \ang{i, j}$ and $\N_2^L \ang{p, q}$.  For clarity, we denote by $\fl(i, j, p, q)$ the minimum number of contractions needed to obtain a common contraction of $\N_1^L \ang{i, j}$ and $\N_2^L \ang{p, q})$.  We get
\begin{align*}
    \fb(a, b, c, d) =
    &~|b - a| + |d - c|~+\fc(\N_1^B \ang{a, b}, \N_2^B \ang{c, d} )~+\\
    &\min \begin{cases}
        \fl(u + 1, a - 1, w + 1, c - 1) + \fl(v - 1, b + 1, x - 1, d + 1) \\
        \fl(u + 1, a - 1, x - 1, d + 1) + \fl(v - 1, b + 1, w + 1, c - 1)
    \end{cases}
\end{align*}
The latter minimization occurs because we do not know which sides of the cycles should be put in correspondence.
It only remains to describe how to compute these lateral subnetworks, i.e.\,entries
$\fl(i,j,p,q)$. To start with, if $L(\N_1^L\langle i,j\rangle)\neq L(\N_2^L\langle
p,q\rangle)$, we set $\fl(i,j,p,q)=+\infty$. Then, in the case of equality, if $i>j$
(which implies
$p>q$ given the constraint $L(\N_1^L\langle i,j\rangle)=L(\N_2^L\langle p,q\rangle)$, the
corresponding subnetworks are empty and $\fl(i,j,p,q)=0$.  For $i, j$ on the same side of the $\N_1$ cycle and $p, q$ on the same side of the $\N_2$ cycle, we consider two cases: either there is an edge $k, k+1$ between $i, j$ and a corresponding edge $k', k'+1$ between $p, q$ that can be kept, or not (in the proofs, the extra leaf $\ell_T$ is there to enforce that if $(k, k+1)$ is kept on the $\N_1'$ lateral path, the corresponding edge in $\N_2'$ is on the other lateral path and not elsewhere).  In the former case, the edge splits the computation into two subproblems, and in the latter we must contract everything.  This yields:
\begin{align*}
        \fl(i, j, p, q) = \min 
        \begin{cases}
        \min_{\substack{i \leq k < j \\ p \leq k' < q}} 
        \fl(i, k, p, k') + 
        \fl(k+1, j, k'+1, q), \\
        \fc(\N_1^S \ang{i, j}, \N_2^S \ang{p, q}) + |j - i| + |p - q|
    \end{cases}
\end{align*}

The chain of recurrences can stop here, since the $\N^S$'s are join subnetworks, which can make use of other $\fc$ entries.
Although not trivial, we can argue that a polynomial number of entries is sufficient to compute $\deltamcc(\N_1, \N_2) = \fc(\N_1, \N_2)$.
%As detailed in the full online pre-print, where the proof of the Theorem below is given, a rough analysis of the worst-case complexity of this computation yields $O(n^9)$, with $n=\max(\N_1,\N_2)$.
\medskip
\begin{thm}
    The value of 
    $\deltamcc(\N_1, \N_2)$ between two weakly galled trees without degree-2 nodes can be computed in $O(n^5)$, with $n=\max(|V(\N_1)|,
    |V(\N_2)|)$.
\end{thm}
\begin{proof}
   We first argue the time complexity  to compute $\fc(\N_1, \N_2) = \deltamcc(\N_1, \N_2)$.
   
    We assume a recursive algorithm that starts with a request to calculate
    $\fc(\N_1, \N_2)$, and computes only smaller entries when they are required
    by our recurrences.
    We store in memory any $\fc$, $\fp$ entry and $\fl$ that is computed for future reuse (memoization). 
    Note that $\fb$ has a slightly different status: it is a mere calculus
    intermediate between $\fp$ and $\fl$.
    Thanks to memoization,
    when $\fc, \fp$, or $\fl$ entries require other entries, we assume they can access their value in
    time $O(1)$.  
    The set of all possible composite join subnetworks has exponential size, so it is important to emphasize that we only calculate the necessary $\fc, \fp$, and $\fl$ entries encountered.
    We need to bound the number of such entries to compute through the whole execution, multiplied by the time needed to compute each individual entry.

    Let us start with $\fp$. We argue that $\fp$ is called on $O(n^2)$ entries, each requiring at most time $O(n^2)$
    to compute. To start with, note that $\fp$ is only called by $\fc$ on pairs of matching sub-networks,
    which in particular have the same set of leaves. The entries of $\fp$ are of four types, which we argue
    separately:
    \begin{itemize}
        \item \textbf{Case 1} -
        $\fp(\N_1 \ang{\dang{u}}, \N_2 \ang{\dang{v}})$: by equality of the sets of leaves, $D(u)=D(v)$.
        By Lemma~\ref{lemma:clade_unicity}, there are at most $2$ nodes $v$ verifying this constraint
        for a given $u$. Since the number of $u$'s is at most the number of nodes of $\N_1$, and since each $u$ can have at most 2 corresponding $v$'s, there are therefore $O(n)$ of these entries. Each of them takes $O(1)$ to compute
        (only need to fetch the required $\fc$ value).

        \item \textbf{Case 2} -
        $\fp(\N_1 \ang{\dang{u}}, \N_2 \ang{C, v, w})$: here, $u,v,w$ must verify $D(u)=D(v+1)\cup D(w-1)$.
        Note that we may have $v+1\neq w-1$ (in which case they form a 2-clade pair) or $v+1=w-1$ (in which
        case they are both the reticulation of $C$). In any case, by Lemma~\ref{lemma:clade_unicity},
        given $u$, there is only $O(1)$ possible $v,w$ verifying $D(u)=D(v+1)\cup D(w-1)$. We thus note that even though $C$ needs to be specified in this type of entry, it is determined by $u$.  This yields $O(n)$
        possible entries, each of them being computed in $(1)$ time.
        \item \textbf{Case 3} - 
        $\fp(\N_1 \ang{C, u, v}, \N_2 \ang{\dang{w}})$: takes time $O(n)$ by a symmetric argument.

        \item \textbf{Case 4} -
        $\fp(\N_1 \ang{C_1, u,v}, \N_2 \ang{C_2, w, x})$: since we must have $D(u+1)\cup D(v-1) = D(w+1)\cup D(x-1)$,
        as above, there is only $O(1)$ possibilities for $w,x$ given $u,v$. This gives $O(n^2)$ possible entries.
        For each of them, the first $4$ cases of the minimization only require time $O(1)$.
        As for the last case, the constraint $D(a)\cup D(b) = D(c)\cup D(d)$
        restricts (by Lemma~\ref{lemma:clade_unicity}) $c,d$ to one possibility given $a,b$. Therefore,
        for each of the $O(n^2)$ entries to $\fp$, the last case of the minimization takes time $O(n^2)$ to compute.
    \end{itemize}
    Overall, the complexity of computing $\fp$ is dominated by Case 4, with an $O(n^4)$ complexity.

    We continue with $\fl$. Let us recall that $\fl(i,j,p,q)$ is defined as 
    $\deltamcc(\N_1^L\ang{i,j},\N_2^L\ang{p,q})$. We call an entry to $\fl$ \emph{valid} if 
    $L(\N_1^L\ang{i,j})=L(\N_2^L\ang{p,q})$. We argue that given $i,j$, there is only
    one pair $p,q$ such that $i,j,p,q$ is a valid entry to $\fl$. Indeed, let $x,y$
    be a pair of nodes of $\N_2$ such that $L(\N_2^L\ang{p,q})=L(\N_2^L\ang{x,y})$. 
    We show that, necessarily, $p=x$ and $q=y$. 

    Let us call $C$ the cycle to which $p,q$ belong, $r$ its root, and $t$ its reticulation.
    Likewise, $C'$ is the cycle containing $x,y$, and $r',t'$ are its root and reticulation.
    Note that $p,q$ and $x,y$ are all internal nodes in their respective cycles 
    (potentially with $p=q$ or $x=y$). As internal nodes, and because $\N_1$ and $\N_2$
    are assumed not to have degree-2 nodes, $p,q$ reach leaves (through an edge
    outside of $C$) that the reticulation $t$ cannot reach. Likewise $t$ reaches
    leaves that children of $p,q$ outside of $C$ cannot reach. Note that these
    leaves are not in $L(\N_2^L\ang{p,q}) = D(p)\setminus D(q+1)$

    Suppose $C\neq C'$, and let us first examine the case where $r'$ is reachable from $p$.

    %If $r'$ descends from $q + 1$, then any leaf reached by $r'$ is not in $\N_2^L \ang{p,q}$ since we exclude $D(q+1)$, and the sets of leaves cannot be equal.  
    If $r'$ descends from $p + 1$, then $p$ has a child outside $C$ that reaches leaves not reached by $r'$, and thus the sets of leaves cannot be equal.  If $p = r'$, then $x, y$ are on a cycle rooted at $p$, and $D(x) \setminus D(y+1)$ excludes leaves reached by $p$ that must be in $D(p) \setminus D(q+1)$.
    Finally, if $r'$ descends from a child of $p$ outside of $C$, the same argument applies.
    % The nodes reachable from $p$ are those reachable from $t$, and those reachable from an internal
    % node of $C$ below $p$, through an edge outside of $C$. In the former case, there is at least
    % one leaf reachable from $t'$ that is not in $D(x)\setminus D(y+1)$, but yet is in $D(p)$,
    % which is a contradiction. In the latter case, $D(x)\subseteq D(t)$, which
    % is entirely excluded from $D(p)\setminus D(q+1)$, yielding again a contradiction.
    
    Therefore $r'$ is not reachable
    from $p$, and thus not reachable from $q$. Note also that if $t$, $x$ or $y$ are reachable
    from $p$, then $r'$ must be reachable from $p$, which we just showed cannot happen.

    In particular $x$ and $p$ must be incomparable. However, they have at least a leaf $\ell$ that they
    can both reach. Consider a path $P$ from $p$ to $\ell$ an a path $P'$ from $x$ to $\ell$,
    and the first node $s$ of $P$ that belongs to $P\cap P'$. Such a node exists, as
    the parent of $\ell$ is on both $P$ and $P'$. In addition, it must be a reticulation (as otherwise 
    its only parent would be an earlier node of $P$ in $P\cap P'$). Likewise, consider the paths
    from the root to $x$ and $p$, and the last node belonging to both. This node is the root of a cycle,
    whose reticulation is $s$, containing both $x$ and $p$. This is in contradiction with $C\neq C'$ in weakly-galled
    networks. 

    Therefore $C=C'$. If $x,y$ is not on the same side as $p,q$, then $x$ reaches some leaves $p,q$ cannot reach,
    which is a contradiction. Therefore $x,y$ and $p,q$ are on the same side of $C$.
    If $p$ is strictly above $x$, it can reach leaves that $x$ cannot, therefore $p=x$. Likewise, 
    if $q$ is strictly above $y$, too many leaves are excluded. Therefore $p=x$ and $q=y$

    Since $i, j$ determine $p, q$, this gives a $O(n^2)$ number of valid entries to $\fl$. 
    
    However, some of the calls
    from $\fp$ to $\fl$ might not be valid. 
    To recognize valid entries in time $O(1)$,
    we iterate as a pre-processing step over the $O(n^4)$ possible entries of $\fl$,
    check for each of them in $O(n)$ whether it is valid or not by comparing reachable leaf sets ($O(n^5)$
    total complexity), and store the result.
    The total time spent in $\fl$ is then $O(1)$ for each invalid entry, and $O(n^2)$
    for each valid entry, per the iteration over $k,k'$. A rough upper
    bound on the number of invalid entries is $O(n^4)$, and we have argued an $O(n^2)$
    upper bound on the number of valid entries, yielding $O(n^4)$ in total. 
    The worst-case complexity of computing $\fl$ entries is therefore $O(n^5)$,
    dominated by the pre-processing.

    To finish, we analyze the complexity of computing $\fc$ entries.
    The first step of $\fc$ is to compare (in $O(n)$) the leaf-sets of the two input networks, to check the validity of the entry.
    We argue that there are $O(n^2)$ valid entries to $\fc$. To see this, 
    note that $\fc$ is only called by $\fp$ and $\fl$ (in addition to the original
    main call). The first 3 cases of $\fp$, and $\fl$ call $\fc$ a constant
    number of times, and only when they are handling a valid entry. As analyzed before, 
    there are $O(n^2)$ of these calls. The only case yielding more calls to
    $\fc$ is the last minimization case of Case 4 of $\fp$, where (through the intermediate $\fb$), $O(n^4)$ calls are given to $\fc$. However,
    these are of the form $\fc(\N_1^B\ang{a,b},\N_2^B\ang{c,d})$. These calls
    are only valid if $D(a)\cup D(b) = D(c)\cup D(d)$. As argued for $\fp$,
    and thanks to Lemma~\ref{lemma:clade_unicity}, there is only one possibility
    for $c,d$ given $a,b$ and this constraint. 
    There are therefore $O(n^4)$ calls to $\fc$, with $O(n)$ spent on each of them
    to check the validity. However, only $O(n^2)$ of them are valid and pass this step.
    It remains to account for the complexity of applying Rules 1\&2 exhaustively,
    which we argue can be done in $O(n^3)$ for each entry (yielding $O(n^5)$
    in total).

    To begin with, we assume that as a first step, we compute and store the
    $D(u)$ sets of both networks $\N'_1$ and $\N'_2$.  This can be done naively
    in time $O(n^3)$ by taking, for each node $u$, the union of the $D(v)$ sets
    for the children $v$ of $u$ (so, $n$ nodes, each requiring $O(n)$ unions in
    time $O(n)$ each, yielding cubic time).
    Then, we assume that these are stored in a data structure that allows
    querying for the existence of a 1-clade or 2-clade in time $O(n)$.  This can
    be achieved in multiple ways, for instance storing them in a hash table
    (which takes time $O(n)$ to hash one set of leaves), or by bucket-sorting
    the elements of the clades by their leaf identifiers, then interpreting them
    as strings and storing them in a trie in which each node contains $O(n)$
    pointers, which can be done without adding to the $O(n^3)$ complexity.

    Let us now analyze the complexity of checking whether Rule 1 or 2 applies.   
    There are $O(n)$ 1-clade nodes that
    are non-reticulation children of $r_1$, and also $O(n)$ 2-clades that contain a child $u$ of
    $r_1$ (to see this, note that $u$ is internal to at most one cycle, and if
    $u, v$ is a 2-clade pair, then $v$ occurs in at most two 2-clade pairs to
    evaluate, the other possibly occurring for the other child of $r_1$ in the
    cycle).  For every such 1-clade or 2-clade, we can 
    query our stored sets of 1-clades and 2-clades
    of $\N'_2$ to see if it is present, each query taking time $O(n)$.  Thus, checking whether one rule should be
    applied takes time $O(n^2)$.  
    Since we may apply
    $O(n)$ rules before reaching matching join subnetworks, we  spend time up
    to $O(n^3)$ to reduce the networks.

    To conclude, we spend $O(n^4)$ computing $\fp$, 
    $O(n^5)$ computing $\fl$ (due to the pre-processing) and $O(n^5)$
    computing $\fc$.  The two latter give the overall worst-case complexity
    of $O(n^5)$.
    
    We next argue that the recurrences are correct.
    Let $\N_1', \N_2'$ be join subnetworks of $\N_1$ and $\N_2$, respectively.  We show by induction on $|V(\N'_1)| + |V(\N'_2)|$ that the recurrences given above are correct.  To this end, denote by $OPT := OPT(\N'_1, \N'_2)$ the minimum number of contractions needed to obtain a common contraction $\M$ of $\N'_1$ and $\N'_2$.  We show that $\fc(\N'_1, \N'_2) = OPT$ and, in the case that $\N'_1, \N'_2$ are prime, that $\fp(\N'_1, \N'_2) = OPT$ also holds.

    The base cases are trivial to verify, so we assume that they do not cover $\N'_1, \N'_2$.  So as an inductive hypothesis, we assume that $OPT(\N''_1, \N''_2) = \fc(\N''_1, \N''_2)$ for any pair of join subnetworks of size smaller than $\N'_1, \N'_2$ (with the same hypothesis for $\fp$ when the networks are prime).
    
    Suppose first that one of the subnetworks is composite.  If Rule 1 or Rule 2 applies to either network, we contract $k$ edges incident to either roots until no such rule applies, resulting in join subnetworks $\N''_1, \N''_2$ by Lemma~\ref{lem:contr-maintains-join}.  Since these rules are safe by Lemma~\ref{lem:rules-are-safe}, these $k$ contractions are also required to obtain $OPT$, implying that $OPT = k + OPT(\N''_1, \N''_2) = k + \fc(\N''_1,\N''_2)$ by induction.  Since this is our value of $\fc(\N'_1, \N'_2)$, this case is verified. 

    Next suppose that no rule applies.  Then by Lemma~\ref{lem:reduced-form}, $\N'_1 = \N_1^1 \ast \ldots \ast \N_1^p, \N_2 = \N_2^1 \ast \ldots \ast \N_2^p$, with $L(\N_1^i) = L(\N_2^i)$.
    It is easily seen that after applying contractions to each pair $\N_1^i, \N_2^i$ separately to make them isomorphic, one obtains a common contraction of $\N'_1, \N'_2$ by identifying the roots of the separate subnetworks.  This justifies
    \[
    \fc(\N'_1, \N'_2) = \sum_{i=1}^p \fp(\N_1^i, \N_2^i) \geq OPT.
    \]
    Moreover, the solution with $OPT$ contractions must, at the very least, apply contractions to make each $\N_1^i, \N_2^i$ subnetwork isomorphic, and, because the subnetworks are edge-disjoint and only intersect at the root,  using induction justifies
    \[
    OPT \geq \sum_{i=1}^p \fp(\N_1^i, \N_2^i) = \fc(\N'_1, \N'_2).
    \]
    This shows that the $\fc$ entries are correct, assuming that all $\fp$ entries are correct.

    Suppose that $\N'_1$ and $\N'_2$ are prime join subnetworks, and consider $\fp(\N'_1, \N'_2)$.
    We handle the same set of cases as in the recurrences.

    \medskip

\noindent 
\textbf{Case 1:}  $\N'_1 = \N_1 \ang{\dang{u}}, \N'_2 = \ang{\dang{v}}$.  In this case, $\N'_1$ consists of a root $r_1$ with a single child $u$, which is the root of a subnetwork of $\N_1$, and $\N'_2$ has a root $r_2$ with a single child $v$, which roots a subnetwork of $\N_2$.  In that case, we may assume that the edges $r_1 \rightarrow u$ and $r_2 \rightarrow v$ are kept, so we may ignore them.  In other words, we can find a sequence of contractions to make $\N_1 \ang{u}$ and $\N_2 \ang{v}$ isomorphic, then attach $r_1$ and $r_2$ to the resulting networks to make them isomorphic.
    Since $OPT$ cannot do better than this, this justifies $\fp(\N_1 \ang{\dang{u}}, \N_2 \ang{\dang{v}}) = \fc(\N_1 \ang{u}, \N_2 \ang{v})$.
    Importantly, we observe that $\N_1 \ang{u}, \N_2 \ang{v}$ are join subnetworks of $\N_1, \N_2$, so the corresponding $\fc$ entry used in our recurrences is well-defined.

    \medskip

\noindent 
\textbf{Case 2:}  $\N'_1 = \N_1 \ang{\dang{u}}, \N'_2 = \N_2 \ang{C, v, w}$.  Here, $\N'_1$ starts with a single edge $r_1 \rightarrow u$,
    while $\N'_2$ starts with a node $r_2$ that is the root of a cycle, which is the fusion of $v$ and $w$ on cycle $C$. 
    
    Suppose that edge $r_1 \rightarrow u$ is conserved in the $OPT$ scenario.
    Then $\N'_2$ must be modified so that its root $r_2$ becomes a node with a single child $x$.  
    This means that the edges of $C$ are contracted so that all children of the cycle nodes that are not on $C$ become a child of $x$.  
    After applying these contractions in the $OPT$ scenario, $r_2$ has a single child, the latter which roots the subnetwork $\N_2^B \ang{C, v+1, w-1}$, which is a join subnetwork.  In this case, we get that $OPT$ is equal to the first case of our recurrence.  

    So suppose that $r_1 \rightarrow u$ is not conserved.  Then after applying this contraction in the $OPT$ scenario, by induction we have $OPT = \fc(\N_1 \ang{u}, \N_2 \ang{C, v, w}) + 1$.  Noting that this $\fc$ entry is well-defined, in this case, $OPT$ is equal to the second case of our recurrence.  

    Since the recurrence takes the minimum of both cases, we get 
    $\fp(\N_1 \ang{u}, \N_2 \ang{C, v, w}) \leq OPT$.  Moreover, it is not hard to see that both cases of the recurrence either yield infinity, or result in a valid common contraction, which justify $\fp(\N_1 \ang{u}, \N_2 \ang{C, v, w}) \geq OPT$.

    \medskip

\noindent 
\textbf{Case 3:} is symmetric to case 2.

    \medskip

\noindent 
\textbf{Case 4:} $\N'_1 = \N_1 \ang{C_1, u, v}, \N'_2 = \N_2 \ang{C_2, w, x}$.  Let $t_1, t_2$ be the respective reticulations of $C_1, C_2$.  The purpose of the lengthy arguments that follow is to establish that $\fc(\N'_1, \N'_2) \leq OPT$.

    Suppose that the $OPT$ scenario contracts an edge incident to $u, v, w,$ or $x$, say $(u, u+1)$ (the other cases are identical).  
    Note that $u + 1$ could not be a reticulation, as such a contraction would be forbidden.
    After contracting this edge in $\N_1 \ang{C_1, u, v}$, we obtain the network $\N_1 \ang{C_1, u, v} / (u, u+1)$, which is a join subnetwork by Lemma~\ref{lem:contr-maintains-join}.
    We therefore get by induction that $OPT = 1 + \fc(\N_1 \ang{C_1, u, v} / (u, u+1), \N_2 \ang{C_2, w, x})$ as in our recurrences.

    So, suppose that in the $OPT$ scenario, no edge incident to $u, v, w, x$ is contracted.
    If we let $W = \{W_1, \ldots, W_l\}$ be a partition of $\N_1 \ang{C_1, u, v}$ that is a witness to the common contraction yielding $OPT$, and $Z = \{Z_1, \ldots, Z_l\}$ the witness partition of $\N_2 \ang{C_2, w, x}$.  By our assumption, $u + 1$ and $v - 1$ are not in the same set as the root of the first network, and $w + 1, x - 1$ are not in the same set as the root of the second network.  This also means that the respective reticulation $t_1, t_2$ of $C_1, C_2$ are also in sets that differ from the roots.
    Let $a, b$ be nodes of $C_1$ such that are in the same witness set as $t_1$, and such that $a - 1$ and $b + 1$ are not in this set (such $a, b$ exist by the previous sentences).
    Define $c, d$ as the analogous nodes of $C_2$, with respect to the set with $t_2$.  Note that any of $a$ and/or $b$ could be equal to $t_1$, and $c$ and/or $d$ could be equal to $t_2$.

    We get that in the $OPT$ scenario,  all the nodes on the path from $a$ to $b$, and from $c$ to $d$, following our cyclic orderings, are contracted.  
    Denote by $\N_1''$ and $\N_2''$ networks obtained after applying these contractions.
    Then $OPT = |b - a| + |d - c| + OPT(\N_1'', \N_2'')$.  
    Unfortunately, we cannot use induction by relating the latter term to an $\fc$ entry, since it might not be a join subnetwork as it applies contractions not at the root.  

    We divide the argument into two subcases: either $C_1$ was contracted to a single edge, or not.  We will assume for now that the entries $\fl(i,j,p,q)$ correctly store the minimum number of contractions needed to obtain a common contraction of the corresponding lateral subnetworks, and we shall prove this later on.

    \medskip 

    \noindent
    \textbf{Subcase 1: $C_1$ contracted to a single edge.}
    Suppose first that $\N_1''$ consists of the root $r_1$ with a single child, i.e.\,$C_1$ got contracted into a single edge.  This is only possible if $a = u + 1, b = v - 1$, and if $C_2$ is also contracted to a single edge, and thus that $c = w + 1, d = x - 1$.
    In that case, we may ignore the dangling root edges, and the optimal contraction only compares the subnetworks rooted at the nodes obtained from the contraction from $a$ to $b$ and from $c$ to $d$.  
    These subnetworks are $\N_1^B \ang{a, b}$ and $\N_2^B \ang{c,d}$, so that
    $OPT(\N_1'', \N_2'') = OPT(\N_1^B \ang{a, b}, \N_2^B \ang{c,d})$.  Because these subnetworks are join subnetworks, this is equal to $\fc(\N_1^B \ang{a, b}, \N_2^B \ang{c,d})$ by induction.
    In other words, $OPT = |b - a| + |d - c| + \fc(\N_1^B \ang{a, b}, \N_2^B \ang{c,d})$.
    Note that these terms are considered in the $\fb(a,b,c,d)$ entry of our recurrence when minimizing over $a, b, c, d$, and we would like to argue that $\fb(a, b, c, d) \leq OPT$.  However, the $\fb$ entry also adds a term from the $\fl$ entries, so let us focus on the latter.

    The considered entries $\fb(i,j,p,q)$ have $i = u+1, j = a-1$ or $i = v-1, j = b - 1$.  Either way, we are in the case where $a = u + 1, b = v - 1$, and $j$ does not descend from $i$.  According to our definition of the $\fl$ terms, they all evaluate to $0$ in $\fb(a,b,c,d)$.  It follows that according to the recurrences, 
    \[
    \fc(\N'_1, \N'_2) \leq \fb(a,b,c,d) = |b - a| + |d - c| + \fc(\N_1^B \ang{a, b}, \N_2^B \ang{c,d}) + 0 = OPT 
    \]

    If $\N''_2$ consists of a root with a single child, a symmetric argument yields the same conclusion.

    \medskip 

    \noindent
    \textbf{Subcase 2: $C_1$ not contracted to a single edge.}
    Let us next suppose that in $\N''_1$, $C_1$ has not been contracted to a single edge, and that in $\N''_2$, $C_2$ is not contracted to a single edge. 
    Then $\N''_1$ has root $r_1$ with the two children $u + 1, v - 1$, which remain in the optimal contraction, and $\N''_2$ has root $r_2$ with children $w + 1, x - 1$. 
    Moreover, the parents of the respective reticulations after contracting the $a-b$ and $c-d$ paths are $a - 1, b + 1$ and $c - 1, d + 1$.
    This means that in the $OPT$ scenario, we must make $|b - a| + |d - c|$ contractions, then make the bottom subnetworks $\N_1^B \ang{a, b}, \N_2^B \ang{c,d}$ isomorphic, then choose corresponding lateral subnetworks and make those isomorphic as well.  

    Noting that our expression for $\fb(a,b,c,d)$ incorporates all of the above, we would like to argue that 
    \[
    \fb(a,b,c,d) \leq OPT.
    \] 
    The main difficulty is that $\fb$ uses the lateral subnetworks $\N_1^L$ and $\N_2^L$, which have an extra leaf that is not considered by $OPT$ (and, as we shall see later on, this extra leaf is necessary for the recurrences to work properly).
    Consider the minimum common contraction $\M$ in which $C_1$ is not contracted to a single edge.  Consider the witness structures $\W$ and $\W'$ of $\N_1 \ang{C_1, u, v}$ and $\N_2 \ang{C_2, w, x}$ that yield the optimal common contraction $\M$.  For a node $y \in V(\M)$, $W_y$ and $W'_y$ denote the part of $\W$ and $\W'$ that got contracted into $y$. 
    Because we assume that $r_1 \rightarrow u+1$ and $r_1 \rightarrow v-1$ are not contracted, $\M$ has a root $r$ with two children, with $W_r = \{r_1\}$ and $W'_r = \{r_2\}$.
    Also, $r$ is the root of a cycle of $\M$ with reticulation $t$, and we must have $a, b \in W_t$ and $c, d \in W'_t$.

    Let $r = z_0, z_1, z_2, \ldots, z_l = t$, $l \geq 1$, be one of the paths of the cycle of $\M$ rooted at $r$, such that $u+1 \in W_{z_1}$ (such a path must exist).
    We infer that $a-1 \in W_{z_{l-1}}$.  We must then have either $w+1 \in W'_{z_1}, c-1 \in W'_{z_{l-1}}$ or $x-1 \in W'_{z_1}, d+1 \in W'_{z_{l-1}}$.  Assume that the former holds (the latter leads to an identical argument).
    
    Denote by $V_1\ang{u+1, a-1}$ the set of all nodes of $\N_1$ reached by $u+1$ but not $a$, and denote $V_2\ang{w+1, c-1}$ analogously.  Notice that the subnetwork induced by $V_1 \ang{u+1, a-1}$ is the same as $\N_1^L \ang{u+1, a-1}$, except that the latter has the extra leaf child $\ell_T$ attached to $a-1$.
    Because edges $r_1 \rightarrow u+1$ and $a-1 \rightarrow a$ are kept, and because these edges cut $V_1 \ang{u+1,a-1}$ from the rest of the network, any set of $\W$ that contains nodes $V_1 \ang{u+1,a-1}$ only contains nodes from that set $V_1 \ang{u+1,a-1}$ (i.e., no overlap with the exterior).  The same holds for $V_2\ang{w+1, c-1}$.  

    This means that $\{W_i \in \W : W_i \subseteq V_1 \ang{u+1,a-1}\}$ and $\{W'_i \in \W : W'_i \subseteq V_2 \ang{w+1,c-1}\}$ form a witness structure to a common contraction of the subnetworks induced by $V_1 \ang{u+1, a-1}$ and $V_2 \ang{w+1, c-1}$.  Moreover, $a-1 \in W_{z_{l-1}}$ and $c-1 \in W'_{z_{l-1}}$, meaning that both sets contract to the same node $z_{l-1}$.  If we add a new leaf $\ell_T$ that is a child of $a-1, c-1$ in these induced subgraphs, and $\ell_T$ as a child of $z_{l-1}$, we see that that the same $W_i$ and $W'_i$ sets still form a witness structure of these networks.  

    All this setup allows us to conclude that if we use the optimal witness structure yielding $\M$, and restrict its subsets to the internal nodes on the lateral networks 
    $\N_1^L \ang{u+1,a-1}$, $\N_2^L \ang{w+1,c-1}$, we get a witness structure of the latter two networks (even with the extra leaf).  It follows that to make these lateral subnetworks isomorphic, the $OPT$ scenario needs at least as many contractions as for the pair of networks $\N_1^L \ang{u+1,a-1}$, $\N_2^L \ang{w+1,c-1}$.
    The same holds for the lateral networks on the other side of the cycle.  If we assume that the $\fl$ entries correctly store the distances for all the lateral networks, we deduce
    \begin{align*}
        OPT \geq&~|b - a| + |c - d|~+ \\
        &~\fc(\N_1^B \ang{a,b}, \N_2^B \ang{c,d})~+ \\
        &~\fl(u+1,a-1,w+1,c-1) + \fl(v-1,b+1,x-1,d+1)
    \end{align*}
    which is greater or equal to $\fb(a,b,c,d)$.

    We do need to assume that the $\fl$ entries are correct.
    So, we initiate a separate proof by induction specifically for the $\fl(i,j,p,q)$ entries.  We want to show that $\fl(i,j,p,q)$ is equal to the number $OPT_L$ of contractions needed to achieve a common contraction $\M_L$ of $\N_1^L \ang{i,j}$ and $\N_2^L \ang{p,q}$.  The base cases described in the text are trivial.
    
    Suppose that some edge $k \rightarrow k + 1$ on the path between $i$ and $j$ is not contracted.  
    Let us choose the smallest $k$ such that this is the case. It means
    that all nodes from $u$ to $k$ (included) are contracted into a single
    point. We argue that there must exist $k'$ in $[p,q]$ such that 
    the edge $k'\rightarrow k'+1$ in $\N_2^L\ang{p,q}$ is not contracted either, and such that the edge separates the leaves in the same manner as $k \rightarrow k+1$.

    In terms of $\M_L$-witness-structure, all nodes on the $i-k$ path are part
    of the same $W_r$, for some $r\in V(\M_L)$. On the other hand,
    $k+1$ belongs to $W_s$ with $s\neq r$. Let $A$ be the set of leaves
    reached by $i$ but not $k+1$, and $B=L(\N_1^L\ang{i,j})\setminus A$.

    Let $\W'$ be a $\M_L$-witness structure in $\N_2^L\ang{p,q}$.
    To start with, since $i$ is the root of $\N_1^L\ang{i,j}$, $r$ is
    the root of $\M_L$, and the root $p$ of $\N_2^L\ang{p,q}$ must be in $W_r'$, the correspondent of $W_r$ in $\W'$.

    Then, in $\M_L$, all paths from the leaf $\ell_T$ to $r$ must encounter
    $s$, as it is the case in $\N_1^L\ang{i,j}$ that all paths from $\ell_T$
    to $W_r$ go through $W_s$. Therefore, in $\N_2^L\ang{p,q}$,
    as $p\in W_r'$, and $\ell_T$ is a child of $q$, there must be
    some node $\tilde{k}\in W_s'$ on the path from $p$ to $q$. We pick $\tilde{k}$
    as close to $p$ as possible. To finish, we argue that $\tilde{k}-1$
    must be in $W_r'$. Indeed, if it is not, then it belongs to some
    other witness set $W_l$. In $\M_L$, $l$ would separate $r$ from $s$,
    which is not the case.
    To conclude, edge $\tilde{k}-1\rightarrow \tilde{k}$ is kept in $\N_2^L\ang{p,q}$.
    We also see that in $\M_L$, the set of leaves reached by $s$ is $B$, from which it follows that the set of leaves reached by $\tilde{k}$ in $\N_2^L \ang{p,q}$ is also $B$.  Since this is also the set of leaves reached by $k+1$, we get that both edges $k \rightarrow k+1$ and $k' \rightarrow k'+1 = \tilde{k}-1 \rightarrow \tilde{k}$ separate the lateral networks into two other lateral subnetworks with the same sets of leaves.

    From this, it follows
    that $\M_L$ can be obtained by making the lateral networks $\N_1^L \ang{i, k}$ and $\N_2^L \ang{p, k'}$ isomorphic, then the networks $\N_1^L \ang{k+1,j}$ and $\N_2^L \ang{k'+1, q}$ isomorphic (the fact that each of those lateral subnetworks has an extra leaf can be argued to not change the optimality as before).  
    Since our recurrence for $\fl(i,j,p,q)$ considers the sum of the two corresponding $\fl$ entries, we ge that $\fl(i,j,p,q) \leq OPT_L$ in this case.

    If an edge on the $p-q$ path is not contracted, a symmetric argument yields the same result.  
    Finally, if no edge on either $i-j$ or $p-q$ path is conserved, we may assume that they are all contracted, resulting in $\N_1^S$ and $\N_2^S$ subnetworks on both sides, and a number of contractions that is $|j-i| + |p-q| + \fc(\N_1^S \ang{i,j}, \N_2^S \ang{p,q})$ by induction (note that the $\fc$ entry is well-defined since $\N_1^S$ and $\N_2^S$ subnetworks are join subnetowkrs).
    Again, this case is considered in our recurrences and we get $\fl(i,j,p,q) \leq OPT_L$.

    In all cases, we get $\fl(i,j,p,q) \leq OPT_L$.  To finish the argument, it is not too difficult to see inductively that each case in the minimization of the $\fl$ recurrence either yields $+\infty$, or yields an integer that corresponds to a scenario leading to a common contraction.  This gives the complementary bound $\fl(i,j,p,q) \geq OPT_L$.
    
    \end{proof}

\section{Conclusion and discussion}
\label{sec:conclusion}

%\noindent 
%\textbf{Conclusion and Discussion.}
We have generalized the original formulation of the Robinson-Foulds
distance on phylogenetic trees, based on edge contractions and expansions, to
phylogenetic networks. This required novel versions of
contractions/expansions capable of creating
and suppressing cycles.  
%We have shown that the space of rooted phylogenetic
%networks on the same set of leaves is connected under these operations.
% This property allows to naturally define on this space an operational distance, 
% that we called $\dce$, as the minimum number of contractions/expansions required
% to transform a network into another.
Both our measures $\dce$ and $\deltamcc$ connect the whole space of networks on the same leaves.
However, whereas in the case of trees, $\dce$ is equivalent
to finding a maximum common contraction, 
that it is not the case for networks, making both measure require their own separate studies.  
% As we deem the outlining of common
% structure in the comparison of networks particularly relevant to 
% application  use-cases (comparing reconstruction methods), we define
% and study $\deltamcc$, a dissimilarity measure for networks based on
% the size of a maximum common contraction. It verifies all properties of
% a distance except the triangle inequality, making it a \emph{semi-distance}.

% In Section~\ref{sec:hardness}, we have proved that deciding
% whether $\deltamcc$ is lower than a given threshold is NP-hard,
% even when the networks have bounded degree, and the size of the
% common contraction is 
% %of size 
% $\leq 3$. Lifting the bounded
% degree constraint, we also prove hardness with a constant number
% of leaves, and one of the networks being a tree. Last,
% our reductions allow to import lower bounds based on the 
% Exponential-Time Hypothesis (Theorem~\ref{thm:eth}).

% Nonetheless, we provide a polynomial-time algorithm for the case
% of weakly-galled trees. As in the computation
% of the Robinson-Foulds distance, the comparison of the sets of \emph{clades} (with
% a suitable generalization) does play a role. However, a fundamental difference
% is that we may not assume a common clade is present in a maximum common
% contraction.

% \subsection{Open problems and future work}

Our work poses several questions and open problems. 
From a \emph{parameterized complexity} point of view,
our reductions prove \textsf{Para-NP-hardness} (i.e.\,NP-hardness
for a constant value of the parameter) of computing $\deltamcc$ with ``size of the common contraction plus maximum degree''
as a parameter, as well as the number of leaves.
However, other parameters, such as the \emph{level} of the input
networks, their \emph{treewidth}~\cite{marchand2022tree}, their \emph{scanwidth}~\cite{berry2020scanning} as well
as the maximum number of allowed contractions, could be of interest and will be the subject of future work. 
Regarding treewidth,
a starting point could be the $O(|H|^{tw(G)+1}|G|)$ algorithm for determining
if a bounded-degree graph $H$ is a contraction of another graph $G$~\cite{matouvsek1992complexity}.
Although the \emph{bounded-degree} constraint is unlikely to help, the complexity of computing $\deltamcc$ on \emph{binary networks} (in+out degree $\leq 3$) is also open.

\backmatter

%\bmhead{Supplementary information}
%The full online version of this article may be found at
%\url{https://arxiv.org/abs/2405.16713}
%
%If your article has accompanying supplementary file/s please state so here. 
%
%Authors reporting data from electrophoretic gels and blots should supply the full unprocessed scans for key as part of their Supplementary information. This may be requested by the editorial team/s if it is missing.
%
%Please refer to Journal-level guidance for any specific requirements.

%\bmhead{Acknowledgements}

%Acknowledgements are not compulsory. Where included they should be brief. Grant or contribution numbers may be acknowledged.
%
%Please refer to Journal-level guidance for any specific requirements.

\section*{Declarations}

The authors declare no conflict of interest nor competing interests.

\subsection*{Funding}

The authors acknowledge financial support from the NSERC/FRQNT NOVA program (grant \#327657).

\bibliography{biblio}% common bib file
\appendix
\renewcommand{\thefigure}{S\arabic{figure}}
\setcounter{figure}{0}

\end{document}